\pdfoutput=1
\documentclass[a4paper,12pt]{article}

\usepackage{slashed}
\usepackage[affil-sl,auth-sc]{authblk}
\usepackage{amssymb,amsmath,amsbsy}
\usepackage{xfrac}
\usepackage{mathrsfs}
\usepackage{mathbbol}
\usepackage{bm}
\usepackage{appendix}
\usepackage{hyperref}
\usepackage{cancel}
\usepackage[top=1.2in, bottom=1.2in,left=0.9in,includefoot]{geometry}               
\usepackage{graphicx}
\usepackage{cite}
\usepackage{float}
\usepackage{caption}
\usepackage{wasysym}
\usepackage{amsthm}
\usepackage{romannum}

\def\code{{\tt SUSY\_FLAVOR}}

\newcommand{\bea}{\begin{eqnarray}}
\newcommand{\eea}{\end{eqnarray}}

\numberwithin{equation}{section}

\title{\bf Distinguishing between MSSM and NMSSM through $\Delta F=2$ processes.}

\author{Jacky Kumar\footnote{email: \tt{jka@tifr.res.in}} $^1$} 

\author{Michael Paraskevas\footnote{email: \tt{mparask@grads.uoi.gr}} $^2$}

\affil{\small $^1$Department of High Energy Physics, Tata Institute of Fundamental Research, \\ 400 005, Mumbai, India }

\affil{\small $^2$Department of Physics, Division of Theoretical Physics, \\ University of Ioannina,  GR 45110, Greece}

\date{September 1, 2016}

\begin{document}

\maketitle
\pagenumbering{arabic}
\begin{abstract}
We study deviations between MSSM and $Z_3$-invariant NMSSM, with respect to their predictions in $\Delta F=2 $ processes. We find that potentially significant effects arise either from the well known double-penguin diagrams, due to the extra scalar NMSSM states, or from neutralino-gluino box contributions, due to the extended neutralino sector. Both are discussed to be effective in the large $\tan\beta$ regime. Enhanced genuine-NMSSM contributions in double penguins are expected for a light singlet spectrum (CP-even,CP-odd), while the magnitude of box effects is primarily controlled through singlino mixing. The latter is found to be typically subleading (but non-negligible) for $\lambda \lesssim 0.5$, however it can become dominant for $\lambda\sim \mathcal{O}(1)$.  
We also study the low $\tan\beta$ regime, where a distinction between MSSM and NMSSM can come instead due to experimental constraints, acting differently on the allowed parameter space of each model. To this end, we incorporate the LHC Run-I limits  
from $H\rightarrow Z\,Z$, $A\rightarrow h\, Z$ and $H^\pm \rightarrow \tau\, \nu $ 
non-observation along with Higgs observables and set (different) upper bounds for new physics contributions in $\Delta F=2 $ processes. We find that a $\sim 25\%$ contribution in $\Delta M_{s(d)}$ is still possible for MFV models, however such a large effect is nowadays severely constrained for the case of MSSM, due to stronger bounds on the charged Higgs masses.
\end{abstract}

\newpage

\section{Introduction}
Among the various low energy realizations of supersymmetry (susy) that have been put forward over the years, the Minimal Supersymmetric Standard Model (MSSM)\cite{DIMOPOULOS1981150} and its simplest extension with a gauge singlet chiral superfield, commonly referred to as Next-to-MSSM (NMSSM) \cite{Fayet:1974pd,Ellis:1988er}, have received most of the attention. This should come as no surprise, since both models display all attractive properties of supersymmetry including gauge coupling unification, solution to the hierarchy problem, natural dark matter candidates and in addition provide with rich, potentially testable, low energy phenomenology. Each model comes with its own advantages. MSSM carries the minimal field content consistent with observations, which by itself is a very strong motivation. On the other hand, NMSSM has a more extended neutral sector but provides with an elegant dynamical origin for the $\mu$-parameter, associating it with  the susy-breaking scale as required for phenomenological reasons. There are several different variants of 
NMSSM \cite{Panagiotakopoulos:1999ah,Barger:2006rd,Cvetic:1997ky}, however in all our analysis we focus only on the \emph{$Z_3$-invariant} NMSSM\cite{Ellwanger:2009dp}. 

The recently discovered $125$ GeV Higgs with SM-like properties at LHC \cite{Aad:2012tfa,Chatrchyan:2012xdj} has imposed strong but distinct constraints on the low energy parameter space of supersymmetric models. In particular, due to the Higgs mass value, MSSM at low $\tan\beta$ is now considered to survive only through hMSSM scenaria \cite{Djouadi:2013vqa,Djouadi:2015jea,Djouadi:2013uqa} which at the same time require a susy-scale up to $\sim 100\,TeV$. Even at large $\tan\beta$, a close-to-maximal mixing in the top-squark sector of the theory is required in order to obtain  the desired Higgs mass radiatively and without setting the susy-scale very high.  On the other hand, in NMSSM this situation is typically more relaxed, with the Higgs field being able to acquire  a larger tree level mass with a low susy-scale,  for small $\tan\beta$ and a relatively large $\lambda$-parameter \footnote{Other mechanisms to obtain a viable SM-Higgs mass, through singlet-light doublet mixing have been also considered in the literature \cite{Cao:2012fz,Jeong:2014xaa} requiring at the same time certain conditions on the mass scales of the theory to be fulfilled.}\cite{Barbieri:2006bg, Hall:2011aa}. Nevertheless, the large $\tan\beta$ regime is typically MSSM-like (and in some cases even worse), requiring analogous large radiative corrections  provided by the same mechanism.  The previous considerations along with the non-observation of supersymmetric particles and effects in focused susy-searches, push the susy-spectrum to higher energies. It is an interesting question therefore to ask how these bounds translate into constraints on the parameter space of each model.    

 In our study we focus on $\Delta F=2$ processes \cite{Altmannshofer:2007cs,Buras:1998raa,Buras:2000dm,Buras:2002vd,Buras:2003td,Buras:2002wq,Buras:2001mb,Buras:2001ra,Virto:2011yx,Virto:2009wm,Queiroz:2016gif} where strong constraints on the masses and flavour structure of supersymmetric  models are known to arise \cite{Gabbiani:1996hi}. In the first part of our analysis we consider every possible source of deviation between the predictions of MSSM and NMSSM. Such effects are in principle expected from the well-known $\tan\beta$-enhanced double penguins (formally two-loop) due to the extra scalar states of NMSSM. However, there is also another source of deviation, commonly neglected in NMSSM studies, which can arise from certain neutralino-gluino box diagrams \emph{(i.e., crossed boxes)}, due to the extended neutralino sector of the theory. Neutralino-gluino diagrams are typically subleading and such effects are usually screened by pure-MSSM contributions from the gluino-gluino boxes which dominate due to the QCD couplings. Nevertheless, at very large $\tan\beta\gtrsim 50$  this is no longer the case and such NMSSM contributions, associated with the bottom Yukawa coupling and also to large neutralino mass insertions ($\propto\lambda v_u$),  become important. We first explain the theoretical mechanism of this effect and then search for regions in the parameter space where it can give significant contributions. Analogous is the approach on double-penguins but since such effects are already extensively discussed in the literature, our analysis stays at a more qualitative level.

Although for large $\tan\beta$ the $\Delta F=2 $ predictions   can be significantly different, for low $\tan\beta$  a distinction between the two models can come instead from the new bounds set by direct searches at LHC. These shift the masses of non-SM particles to higher scales, suppressing the flavour violating processes in which these particles act as mediators, in a ``decoupling" sense. In the second part of our analysis we consider the LHC Run-I limits for New Physics (NP) from $H\rightarrow Z\,Z$, $A\rightarrow h\, Z$ and $H^\pm \rightarrow \tau\, \nu $ non-observation in Heavy-Higgs searches, along with the Higgs observables. We study their impact in Minimal Flavour Violating (MFV) scenaria\cite{D'Ambrosio:2002ex,Barbieri:2011ci} where large NP contributions from charged-Higgs diagrams have been predicted\cite{Barbieri:2014tja}. Such bounds are found to act differently on the two models and although both can still predict a $\sim 25\%$ effect, the parameter space of MSSM is severely constrained. Moreover, in NMSSM where charged-Higgs masses can be lighter, the maximal NP-effects exceed $\sim 30\%$.

Our study is organized as follows: In sec.\ref{2}, after  setting  conventions, the discussion begins with a qualitative analysis of neutralino-gluino box effects. We study the mechanism that generates enhanced genuine-NMSSM contributions with the help of the Flavour Expansion Theorem\cite{Dedes:2015twa}. Having this analysis as a guide, we proceed to  quantify these effects, calculating $\Delta M_{s(d)}$ in mass eigenstate basis.  The section ends with a study of double-penguin NMSSM contributions, embedding possible effects and behaviours within a common theoretical framework.  In sec.\ref{3}, we follow a different strategy and translate Higgs and Heavy-Higgs data into bounds on the
$m_{H^\pm}-\tan\beta$ planes of MSSM and NMSSM. Subsequently, these bounds are used in order to distinguish between the two models through $\Delta F=2$ observables, in MFV scenaria. We conclude with a summary of our results in sec.\ref{4}.  The technical tools, required for calculations, and a discussion of the NMSSM potential in the parameter space of enhanced effects, namely at large $\tan\beta,\lambda$, are given in the appendices. 

Note also that in all our analysis we use a more generalized  concept of ``flavour", as also used in \cite{Dedes:2015twa}. In brief, we refer to flavour as a \emph{non-trivial internal space of eigenstates which produces mixing effects}. In this sense, there is squark, Higgs, neutralino, etc., flavour space, while there is no such space for gluons or gluinos.

For the plots in sec.\ref{2}, we have used the publicly available \code \cite{Rosiek:2010ug,Crivellin:2012jv,Rosiek:2014sia}, 
MSSM-code to perform the full 
calculation in \emph{mass basis} taking into account all box and Double Penguin contributions. In this, we have implemented the required modifications due to the extended neutralino and Higgs sectors of NMSSM as well as the extra modifications for chirally enhanced effects in NMSSM, as explained in sec.\ref{generaldp}. In all  checks the MSSM limit of NMSSM is correctly reproduced. In addition, we have considered the latest QCD hadronic matrix elements and the re-adapted SM predictions for $\Delta M_{s(d)}$, recently published by Fermilab Lattice and MILC collaborations \cite{Bazavov:2016nty}.

\section{$\Delta F=2$ processes in MSSM and NMSSM}\label{2}
\subsection{General considerations in meson anti-meson mixing}
In terms of Effective Field Theory (EFT) the amplitude for B-meson mixing is defined as
$M_{12}^q = \left < B_q|H_{eff}| \overline{B_q} \right > $, 
where $q=d,s$ stand for $B_d,B_s$ mixing, respectively.
The effective Hamiltonian, $H_{eff}$, can be consistently expressed in the basis of eight dimension-six operators $Q_i$ as,
\begin{equation}
H_{eff} = \sum_{i} C_i Q_i +  \emph{h.c},
\label{eq:heff}
\end{equation}   
with $C_i$ being their respective Wilson Coefficients (WC). We follow the operator basis defined in \cite{Buras:2002vd}, which reads explicitly, 
\begin{equation}
Q^{VLL} =(\bar b_L \gamma_{\mu} q_L)(\bar b_L \gamma^{\mu}  q_L) \ , \quad  
Q^{VLR}=(\bar b_L \gamma_{\mu}  q_L)(\bar b_R \gamma^{\mu}   q_R) \ , \quad  
Q^{SLR}=(\bar b_R q_L)(\bar b_L q_R)  \ , \quad  \nonumber
\end{equation}
\begin{equation}
Q_1^{SLL}=(\bar b_R q_L)(\bar b_R q_L) \ , \quad  
Q_2^{SLL}=(\bar b_R \sigma_{\mu \nu} q_L)(\bar b_R \sigma^{\mu \nu} q_L) \ , \quad  
\label{bbops}
\end{equation}
\begin{figure}[t]
\centering
  \includegraphics[trim={3cm 18cm 0 3.8cm},clip]{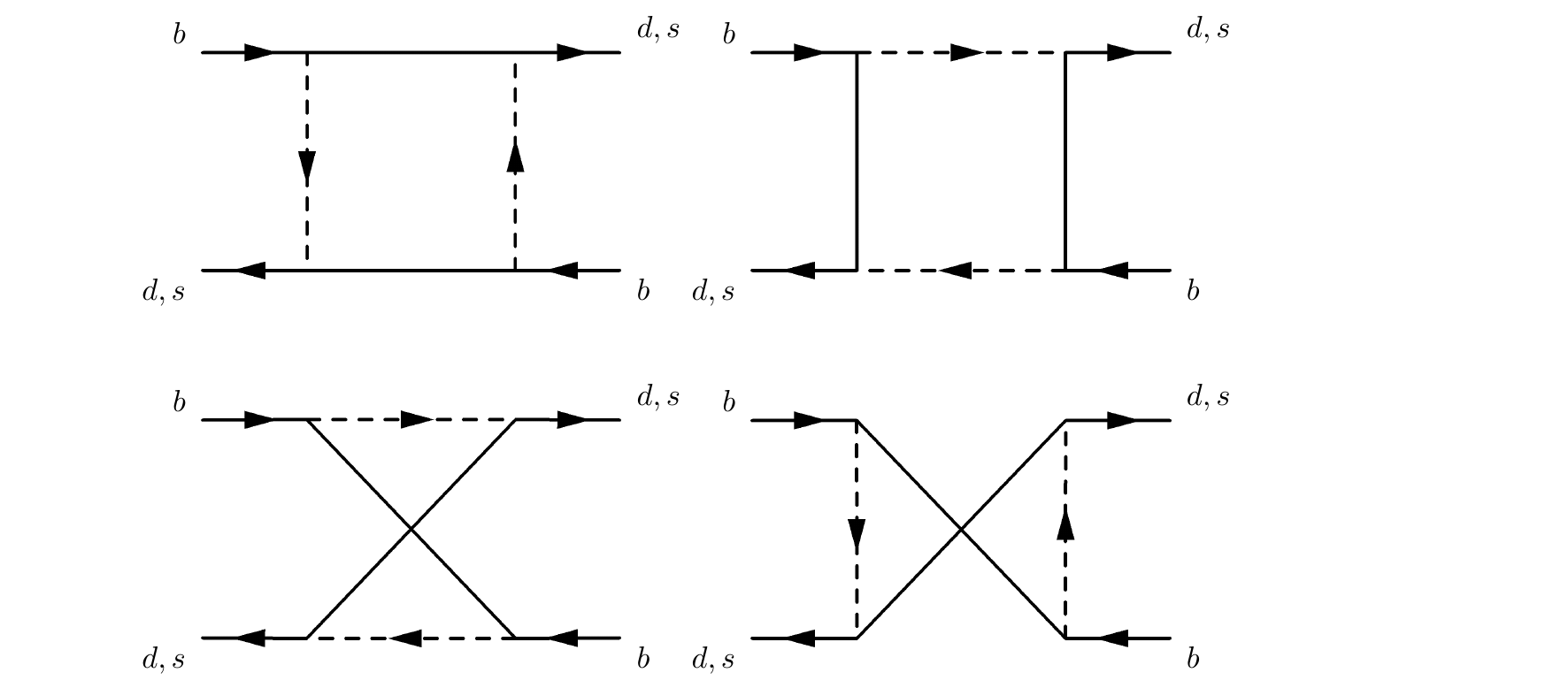}
  \caption{\small One-loop box diagrams contributing to $\Delta F=2$ observables.}
\label{fig:mssm}
\end{figure}\normalsize
In the above expressions, the diagonal quark-color indices are suppressed (assumed to be contracted separately within each bracket), and $\sigma_{\mu \nu} =\frac{1}{2}[\gamma_{\mu},\gamma_{\nu}]$. The remaining three operators, 
$Q^{VRR}, Q_1^{SRR}, Q_2^{SRR}$  are obtained from 
$Q^{VLL}, Q_1^{SLL}, Q_2^{SLL}$ by 
interchanging $L$ with $R$. 
In SM only $Q^{VLL}$ gets non-zero contribution from one-loop \emph{box diagrams} with quarks 
and $W$-bosons circulating in the loops. But in MSSM there are various, additional, box contributions  mediated by:{ \romannum{1}) charged Higgs, up-quarks; \romannum{2}) charginos, up-squarks; \romannum{3}) gluinos, down-squarks; \romannum{4}) neutralinos, down-squarks; \romannum{5}) gluino, neutralino, down squarks }\cite{Altmannshofer:2007cs}. Their diagrammatic topologies are shown in Fig.\ref{fig:mssm}. Certain two-loop diagrams (\emph{i.e., double-penguins}) which depend on positive powers of $\tan\beta$ become also relevant for large values of this parameter and can easily dominate over any other contribution \cite{Buras:2002vd}.

In NMSSM, the situation with respect to $\Delta F=2$ effects is quite similar but not identical. Genuine-NMSSM contributions are either related to box diagrams involving neutralinos or double penguins. For the former, this is understood from the presence of an extra neutralino state in box interactions. Although it affects rather trivially the neutralino - down quark - down squark vertex and the summations over internal neutralino states, in some cases it can leave a strong imprint on the observables. Such contributions can in principle arise from diagrams of type (\romannum{4}) or (\romannum{5}), however only the latter will be shown to be important. For double penguins the situation is more involved and although the effect of an extra neutralino circulating in loops is in practice irrelevant, the extra CP-even and CP-odd singlet states induce various modifications in relevant couplings and spectra. 

All such details are discussed in what follows, where we isolate the origin of \emph{genuine-NMSSM} effects in box and double penguin contributions, without taking into account any \emph{a priori} assumption on spectrum or flavour structure of the theory. This eventually leads us to regions of the parameter space where these contributions are expected to be enhanced and a subsequent numerical analysis quantifies their magnitude and verifies our arguments. On equal footage, our study suggests that any other region of the parameter space is expected to give the same predictions in MSSM and NMSSM with respect to $\Delta F=2$ processes,  as long as corresponding parameters are allowed in both models.  

\subsection{NMSSM contributions in box diagrams}
\subsubsection{Genuine-NMSSM contributions from neutralinos}

The scale invariant superpotential of $Z_3$-NMSSM in the presence of the  singlet superfield $\hat S $, reads,
\begin{equation}
{   W_{NMSSM}} =    W_{MSSM}\Big|_{\mu=0} + \lambda \,\hat S \, \hat H_u
\hat H_d + {\kappa\over 3}~  \hat  S^3,
\label{eq:poteq}
\end{equation}
where an effective $\mu$-parameter is generated when the singlet scalar $  S$ acquires a non-vanishing vacuum expectation value (vev), as  $\mu_{eff}= \lambda <S>\equiv \lambda {v_s\over \sqrt{2}}$.  Our conventions for $W_{MSSM}$ and soft sector follow those of 
\cite{Rosiek:1989rs,Rosiek:1995kg} to which we add the genuine-NMSSM soft terms 
\begin{equation}
-\mathcal{L}_{\textit{soft}}^N = m_S^2 |S|^2+ (\lambda A_\lambda H_u H_d S + {1\over 3} \kappa A_\kappa S^3 + h.c. )
\end{equation}

First we focus on the neutralino mass matrix of NMSSM, and discuss the modifications that this brings to the theory, as compared to MSSM. The symmetric neutralino mass matrix, in flavour basis $(\tilde{B}, \tilde{W}, \tilde{H_d^0}, \tilde{H_u^0}, \tilde{S})$ is given by  
\begin{equation} 
\bf{M_N}=\left( \begin{array}{ccccc}
M_1 &0 &-\frac{e v_d}{2 c_{\rm w}} & \frac{e v_u}{{2} c_{\rm w}}  & 0  \\
 &M_2   & \frac{e v_d}{2 s_{\rm w}} & -\frac{e v_u}{2 s_{\rm w}} &0 \\
 &  & 0 & - \mu_{eff} &  -{\lambda v_u\over \sqrt{2}}\\
 &  &  & 0 &  - {\lambda v_d\over \sqrt{2}} \\
 &  &  &  &  2 \kappa v_s\over{\sqrt 2} \end{array} \right)  
 \end{equation}
where $M_1$ and $M_2$ are the bino and wino flavour masses, respectively, and for simplicity all parameters are taken to be real. One can easily notice, in the matrix above, that the NMSSM effects are isolated in the extra fifth dimension and the mixing of the genuine-NMSSM state (\emph{i.e., singlino}) with other neutralinos is controlled by the $\lambda$ parameter. In the well-known MSSM limit of NMSSM, namely for $\lambda\sim\kappa\rightarrow 0$ (keeping $\mu_{eff}\neq 0$) this mixing vanishes and genuine NMSSM effects decouple. Conversely, for large values of $\lambda$  singlino mixing increases, giving rise to enhanced effects.

In order to show how the singlino state affects $\Delta F=2$ observables, we employ the couplings relevant to neutralino-related box diagrams. Neutralino-down quark-down squark vertices read,
\begin{align}
 &\big(V_{\chi D d}^{L}\big)_{Iia} = -\frac{e}{\sqrt 2 ~s_{  \textrm{w}} c_{\textrm{w}} } ~(Z_{D})_{Ii} \left(  \frac{s_{  \textrm{w}}}{3} (Z_N)_{1a}  - c_{  \textrm{w}} ~(Z_N)_{2a}\right) + Y_d^I (Z_D)_{I+3,i} ~(Z_N)_{3a} 
\label{VL} \\
& \big(V_{\chi D d}^{R}\big)_{Iia} = -\frac{e\sqrt{2}}{3c_{  \textrm{w}}} (Z_D)_{I+3,i} (Z_N)_{1,a}^{*} +  Y_d^I (Z_D)_{Ii} (Z_N)_{3a}^{*}\label{VR}
\end{align}
where $Z_D$, $Z_N$ are the rotation matrices of down squarks and neutralinos respectively. Such couplings have the same form in MSSM and NMSSM, but the neutralino index $a$ for the latter case runs up to 5 (instead of 4).  

The first important thing to notice is the fixed indices of the $Z_N$ rotation matrices. Since in flavour basis there is no down quark - down squark  - singlino (or $\tilde H_u^0$ ) coupling,  this property is inherited to the rotation matrices of the mass basis whose external (fixed) indices range from 1 to 3. As a result, singlino effects can only arise through mixing with these states, namely with the higgsino $\tilde H_d^0$ and the gauginos.   

Another thing to notice is the presence of down-type Yukawa couplings in the higgsino part of the vertices in eqs.(\ref{VL}-\ref{VR})(\emph{i.e.,} $(Z_N)_{3a}^{(*)}$ always comes together with $Y_d^I$). For low $\tan\beta$, higgsino contributions are small due to these couplings, even as compared to gauginos. However for large $\tan\beta$, they become significantly enhanced, especially if they involve only the bottom Yukawa coupling $Y_b$ which in this case is comparable to the strong QCD-coupling. 

\begin{figure}[t]
\centering
  \includegraphics[trim={3cm 22cm 0 3.8cm},clip]{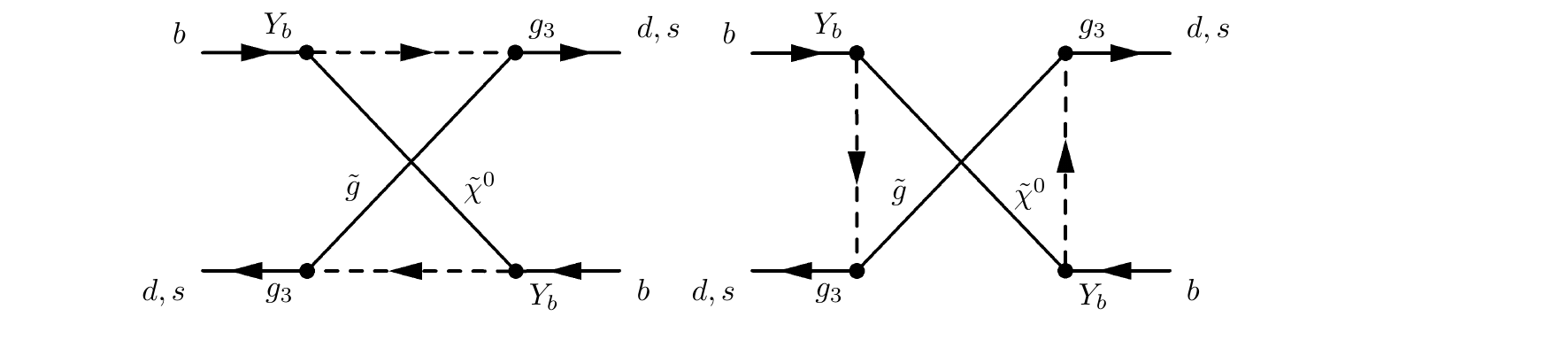}
  \caption{\small Neutralino-gluino box contributions in mass basis, mediating genuine-NMSSM contributions proportional to $g_3^2 Y_b^2$, which become enhanced in the large $\tan\beta$ regime.}
\label{fig:nmssm}
\end{figure}

At this point it is instructive to search for higgsino contributions which are associated only with $Y_b$ and therefore mediate the leading effects at large $\tan\beta$. As shown in Fig.\ref{fig:nmssm}, only certain neutralino-gluino crossed boxes carry this property. The neutralino-neutralino boxes will always involve $Y_{d}^{1,2}$ (or gaugino couplings) instead of $g_3$ and therefore will be subleading. Thus, we may safely neglect them in our analysis. Analogous is the argument for \emph{Kaon-mixing} which involves only Yukawas of the first two generations. Hence, we focus on $B_{s(d)}$ mixing where all neutralino-gluino box contributions in WC are given in \ref{app:ngwc} (mass basis). There, the leading higgsino crossed-box contributions ($\propto g_3^2Y_b^2$) are understood as terms which involve only the higgsino part of the vertices $\big(V_{\chi D d}^{L}\big)^{(*)}_{{3}ia},\big(V_{\chi D d}^{R}\big)^{(*)}_{{3}ia}$. To isolate these effects from gauginos, one can use an approximation on the vertices \eqref{VL},\eqref{VR},  which reads,
\begin{align}
 &\big(V_{\chi D d}^{L}\big)_{3ia} \approx  Y_b (Z_D)_{6i} (Z_N)_{3a} 
\label{VLh} \\
 &\big(V_{\chi D d}^{R}\big)_{3ia} \approx  Y_b (Z_D)_{3i} (Z_N)_{3a}^{*}\label{VRh},
\end{align}
 and which becomes effective in the enhancement (large $\tan\beta$) region.

The neutralino mass and squared mass matrices, both relevant for neutralino mixing effects also simplify for $v_d\ll v_u$ giving,   
\begin{align} 
\bf{M_N}\approx\left( \begin{array}{ccccc}
M_1 &0 &0 & \frac{e v_u}{{2} c_{\rm w}}  & 0  \\
 &M_2   & 0 & -\frac{e v_u}{{2} s_{\rm w}} &0 \\
 &  & 0 & - \mu_{eff} &  -{\lambda v_u\over \sqrt{2}}\\
 &  &  & 0 &  0 \\
 &  &  &  &  2 \kappa v_s\over{\sqrt 2} \end{array} \right), ~
\bf{M_N^2}\approx\left( \begin{array}{ccccc}
M_{11}^2  &M_{12}^2 & M_{13}^2 & M_{14}^2  & 0  \\
 &M_{22}^2   & M_{23}^2 & M_{24}^2 &0 \\
 &  & M_{33}^2 &  0 &M_{35}^2 \\
 &  &  &M_{44}^2 & M_{45}^2   \\
 &  &  &  &  M_{55}^2 \end{array} \right) \label{eq:MNMN2}
 \end{align}
 \begin{eqnarray}
 &M_{11}^2=M_1^2 + \frac{e^2 v_u^2}{4 c^2_{\rm w}}, ~ M_{22}^2=M_2^2 + \frac{e^2 v_u^2}{4 s^2_{\rm w}} ,\nonumber\\
 & M_{33}^2=\mu_{eff}^2+{\lambda^2 v_u^2\over 2}, ~ M_{44}^2=\mu_{eff}^2+ {e^2v_u^2\over 4 c^2_{\rm w} s^2_{\rm w} },~M_{55}^2=2 \kappa^2 v_s^2+{\lambda^2 v_u^2\over 2}.
 \end{eqnarray}
The off-diagonal entries of $M_N^2$, associated with genuine NMSSM effects, are
\begin{eqnarray}
M_{35}^2 = -({ \kappa v_s} )({\lambda v_u}) , ~
M_{45}^2 = \mu_{eff}\Big({\lambda v_u\over \sqrt{2}}\Big), 
\end{eqnarray} 
while all other off-diagonal entries are pure-MSSM. This is easily understood from the fact that the latter are $\lambda,\kappa,v_s$-independent.

We  now have all tools required to study genuine-NMSSM contributions, arising from the $\tan\beta $ enhanced neutralino-gluino crossed boxes. For this purpose we employ the  Flavour Expansion Theorem (FET) \cite{Dedes:2015twa, Paraskevas:2015eby,Rosiek:2015jua} which makes the transition from mass to flavour basis QFT a trivial process. 

The crossed boxes of Fig.\ref{fig:nmssm} display a fixed neutralino flavour-basis structure. To see this, one first substitutes the vertices \eqref{VLh},\eqref{VRh} in the general expressions of \ref{app:ngwc}. Keeping only relevant terms, the mass-basis expressions display the form,
\begin{eqnarray}
& m_{\tilde g} ~(Z_N)_{3a} ~m_a D_0(m_{\tilde g}^2 , m_a^2, x) ~(Z_N)_{3a}\label{eq:puremass}\\
&(Z_N)_{3a} ~D_2(m_{\tilde g}^2, m_a^2,x) ~(Z_N)_{3a}^*\label{eq:mixedmass}
\end{eqnarray}
where all irrelevant factors are neglected (including $g_3^2Y_b^2$) and the down-squark flavour space (\emph{i.e.,} $Z_D$ and  the two down-squark mass arguments of the loop-functions) is suppressed into the argument $x$ since it is irrelevant to neutralino space. Up to complex conjugation in the above expressions, one can easily verify that there is no other  structure \cite{Dedes:2015twa}. 

Next, by applying FET one directly translates the mass eigenstate expressions, into the corresponding Mass Insertion Approximation (MIA) expansions, which then read,
\small\begin{align}
 m_{\tilde g} ~ \Big[\mathbf{M_N} D_0(m_{\tilde g}^2,\mathbf{M_N^2},x)\Big]_{33}&=
 m_{\tilde g} ~ (M_N)_{35}~(M_N^2)_{53}~ E_0\Big(m_{\tilde g}^2,(M_N^2)_{55} ,(M_N^2)_{33},x \Big)+\dots\label{eq:pureflav}\\
 \Big[D_2(m_{\tilde g}^2, \mathbf{M_N^2},x)\Big]_{33}&= ~D_2\Big(m_{\tilde g}^2, (M_N^2)_{33},x\Big)+\dots \label{eq:mixedflav}
\end{align}\normalsize
where dots represent terms at higher FET-order (\emph{i.e.,} higher order in neutralino mass insertions, $M_N^2$). The explicit form of all relevant loop functions is given in \ref{app:loop}.

\begin{table}[t]
\begin{center}
\small\begin{tabular}{c||cccccccc}
$Q_i$&$Q^{VLL}$&$Q^{VRR}$&$Q_1^{SLL}$&$Q_1^{SRR}$
&$Q_2^{SLL}$&$Q_2^{SRR}$& $Q^{VLR}$  & $Q^{SLR}$  \\ &&&&&&&\vspace{-0.4cm}\\ 
 \hline  \vspace{-0.3cm}  \\
\small{MIs}& $\delta_{LR}^{q3}\delta_{LR}^{q3}$
&  $\delta_{RL}^{q3}\delta_{RL}^{q3}$
&  $\delta_{LL}^{q3}\delta_{LL}^{q3}$
&  $\delta_{RR}^{q3}\delta_{RR}^{q3}$
&  $\delta_{LL}^{q3}\delta_{LL}^{q3}$
&  $\delta_{RR}^{q3}\delta_{RR}^{q3}$
&  $\delta_{LL}^{q3}\delta_{RR}^{q3}$
&  $\delta_{LL}^{q3}\delta_{RR}^{q3}$  \\
&&&&&&&\vspace{-0.35cm}\\
&&& &  &  &  &  $\delta_{RL}^{q3}\delta_{LR}^{q3}$
&$\delta_{RL}^{q3}\delta_{LR}^{q3}$ \\
&&&&&&&\vspace{-0.35cm}\\
\hline \vspace{-0.3cm}\\
\small{NMSSM} &\emph{genuine} 
&\emph{genuine}  &\emph{genuine}  &\emph{genuine}  &\emph{genuine} &\emph{genuine}  & 
\emph{mixed} &
\emph{mixed}\vspace{-0cm}\\ 
\end{tabular}
\caption{\small Down-squark flavour dependence of \emph{genuine} and \emph{mixed} NMSSM contributions, related to higgsino-singlino crossed boxes. It is obtained by isolating all terms displaying the structure of \eqref{eq:puremass} and \eqref{eq:mixedmass} in the full expressions of \ref{app:ngwc} and subsequently applying the MIA for down-squarks. Here $\small{q}=1,2$ refers to $B_d,B_s$ -mixing respectively.}\label{tab:mi}
\end{center}
\end{table}
\normalsize
The leading \emph{genuine-NMSSM effects} come from $E_0$-terms, having a strong dependence on $\lambda,\kappa$-parameters through $(M_N)_{35}$ and $(M_N^2)_{53}$ which are, in addition, related to $v_u$. Although  suppressed by a neutralino mass insertion they can become important when $\tilde H_d^0 - \tilde S$  mixing is sufficiently large. The $D_2$-terms are less sensitive to the NMSSM parameters $\lambda,\kappa$ since these appear only through the $(M_N^2)_{33}$ argument of the respective loop function. In this sense, $D_2$-terms  mediate \emph{mixed effects} which is understood by the fact that they are non-zero in the MSSM limit, $\lambda\sim\kappa\to 0$.  Typically, the $E_0$-terms are safe from $D_2$-term screening, since they are primarily associated with different types of squark mass insertions, as shown in Table \ref{tab:mi}. Nevertheless, due to neutralino mass insertion suppression, the $E_0$-term can become comparable to other neutralino-gluino MSSM contributions. These are subleading in the couplings (\emph{e.g.,}$\propto Y_b Y_s,g_2^2$,etc.) but not suppressed by neutralino insertions. In the following numerical analysis section, we discuss the relative magnitude of genuine-NMSSM and MSSM contributions, in the $\tan\beta$ enhanced region and with respect to the size of higgsino-singlino mixing. 

Before concluding this qualitative analysis, an important remark should be made for one-loop diagrams that do not involve neutralinos and therefore, by default, mediate pure-MSSM effects. If these dominate in the NMSSM enhanced regions, they can potentially screen neutralino-related contributions altogether, thus making our discussed effect negligible. 
\begin{itemize}
\item We define as \emph{``MSSM-screening"} or simply \emph{``screening"} the general property that some pure-MSSM contribution may be sizeable in the same region of the parameter space, where we study our effects. 
\end{itemize}
If the screening is large, the room for other NP contributions in general and genuine-NMSSM contributions in particular, becomes small. In order to stay within the experimental bounds, one needs therefore to consider smaller squark mass insertions (or larger masses) which results to a suppression of all flavour violation effects and thus to a suppression of genuine-NMSSM effects, alongside. 

In $\Delta F=2$ observables there are various potential sources for MSSM screening. The diagrams involving charged Higgs and charginos are associated with up-quarks and up squarks, respectively. The former carry a fixed SM flavour violation mechanism, manifesting itself through the CKM matrix and up-quark masses and their contribution is independent of squark flavour violation. Thus, their effect is unrelated to any other contribution and they become suppressed for very large values of $M_A$. The latter,  depend on the flavour structure of the up squark sector and carry a CKM-dependence, as well. Only when $(m_{Q}^2)_{ij}$ off-diagonal soft masses are considered, their effect becomes correlated to neutralino diagrams through $\delta_{LL}$ mass insertions. In any other case they are independent. Finally gluino-gluino diagrams are the most important screening effects since, as will be discussed, they are controlled by analogous down-squark insertions and in addition they couple to quarks-squarks with the strong QCD-coupling, $g_3$. 

\subsubsection{Numerical analysis of $\Delta M_{s(d)}$ in the NMSSM-enhanced region}
By taking into account the previous qualitative analysis, one is naturally guided to the NMSSM-enhanced region of parameter space,  considered in our figures. It possesses the following common properties for $\Delta M_{s}$ and $\Delta M_{d}$ :
\begin{itemize}
\item Large values of $\tan\beta$ and $\lambda\sim\kappa$ are required. The former condition enhances the down-type Yukawa couplings  which are present in $\tilde H_d^0$ interactions. The latter condition is required for large higgsino-singlino mixing which controls the size of genuine-NMSSM contributions. Typical values for significant effects are $50\lesssim \tan\beta~ (\lesssim 65)$ and $0.5\lesssim \kappa\sim\lambda~(\lesssim 1) $.
\item Large values of $M_A$ are preferable, which suppress both charged Higgs contributions and double penguins effects. This is also motivated by the Higgs potential in the large $\tan\beta,\lambda$ regime of NMSSM, as discussed in \ref{app:hmin}. There, we display the method of obtaining phenomenologically viable CP-even and CP-odd scalar masses by fitting the soft $A_\lambda, A_\kappa$ parameters (eqs.\eqref{eq:Alam},\eqref{eq:Akap}), while keeping $\lambda,\kappa$ as free parameters. The typical range for $M_A$  obtained this way is $4 ~TeV\lesssim M_A\lesssim 12 ~TeV$, depending on $\mu_{eff},\tan\beta$ inputs.
\item Genuine-NMSSM effects at one-loop originate from neutralino-gluino box diagrams which are comparable to gluino-gluino diagrams with analogous dependence on down-squark mass insertions. Down squark flavour violation essentially acts as a \emph{``common factor"} in both types of contributions. It can enhance or suppress effects altogether, depending on the size of mass insertions considered. We choose soft masses and mass insertions, so that for $m_{\tilde{g}}=1.1\,TeV$ : i) NP contribution from MSSM is  $\sim (-10\%)$ and $\sim (-20\%)$  for $\Delta M_{s}^{NP}$ and $\Delta M_{d}^{NP}$, respectively, as currently favoured by theoretical and experimental considerations\cite{Bazavov:2016nty,Blanke:2016bhf}) ii) The lightest down-squark mass eigenvalue satisfies $m_{\tilde{d}}^{min}>400$ GeV.; iii) Experimental bounds on flavour and other related observables are satisfied. 

In order to set this MSSM-background, a split down-squark spectrum is considered in our plots. The diagonal soft squark masses $(m^2_{Q})_{ii},(m^2_{U})_{ii}$ are taken at the common scale $M_S=3~TeV$, while $(m^2_{D})_{ii}$ is kept at a relatively light scale ($650~GeV$). The choice of $M_S$ is made for pure convenience, mainly in order to always stay ``safe" from other flavour observables which could constrain the NMSSM-enhanced parameter space. In any case, one may vary squark masses in general, within the MSSM and NMSSM physical parameter space or even introduce other small sources of pure-MSSM flavour violation from the up-squark sector. As long as these remain subleading to the neutralino-gluino contributions, they cannot screen the considered genuine-NMSSM effect.
\item The value of $\mu_{eff}$ lies close to the electroweak scale. This is because $\mu_{eff}$ is also related to $\tilde H_d^0-\tilde S$ mixing which mediates the leading genuine-NMSSM contributions, as discussed previously. Large  effects are induced when $\mu_{eff}\sim \lambda v_u$ and $\lambda\sim\kappa$. Due to perturbativity considerations we take here an upper rough bound $\lambda\lesssim 1$, which already requires a UV-completion for NMSSM before the GUT-scale. As a result, the rough constraint $\mu_{eff} \sim v_u $ is imposed, which numerically translates into $\mu_{eff}\lesssim 300\,GeV$ for non-negligible effects to be produced. For  figures the moderate $\mu_{eff}=180 ~GeV$ is considered, but we note that larger effects are induced for even smaller values. We also set a lower bound for the lightest neutralino state (mixed higgsino-singlino) at $m_{\chi^0}^{min}>{M_Z / 2}$ and which is easily satisfied. Finally, we take $M_1=M_2=1\,TeV$ as a reference value, however we find that even for lighter gaugino masses the qualitative characteristics of our discussion are not affected, although gaugino screening is increased.
\end{itemize} 

\begin{figure}[t]
\hspace{-1.5cm}\includegraphics[height=7.5cm,width=10.6cm]{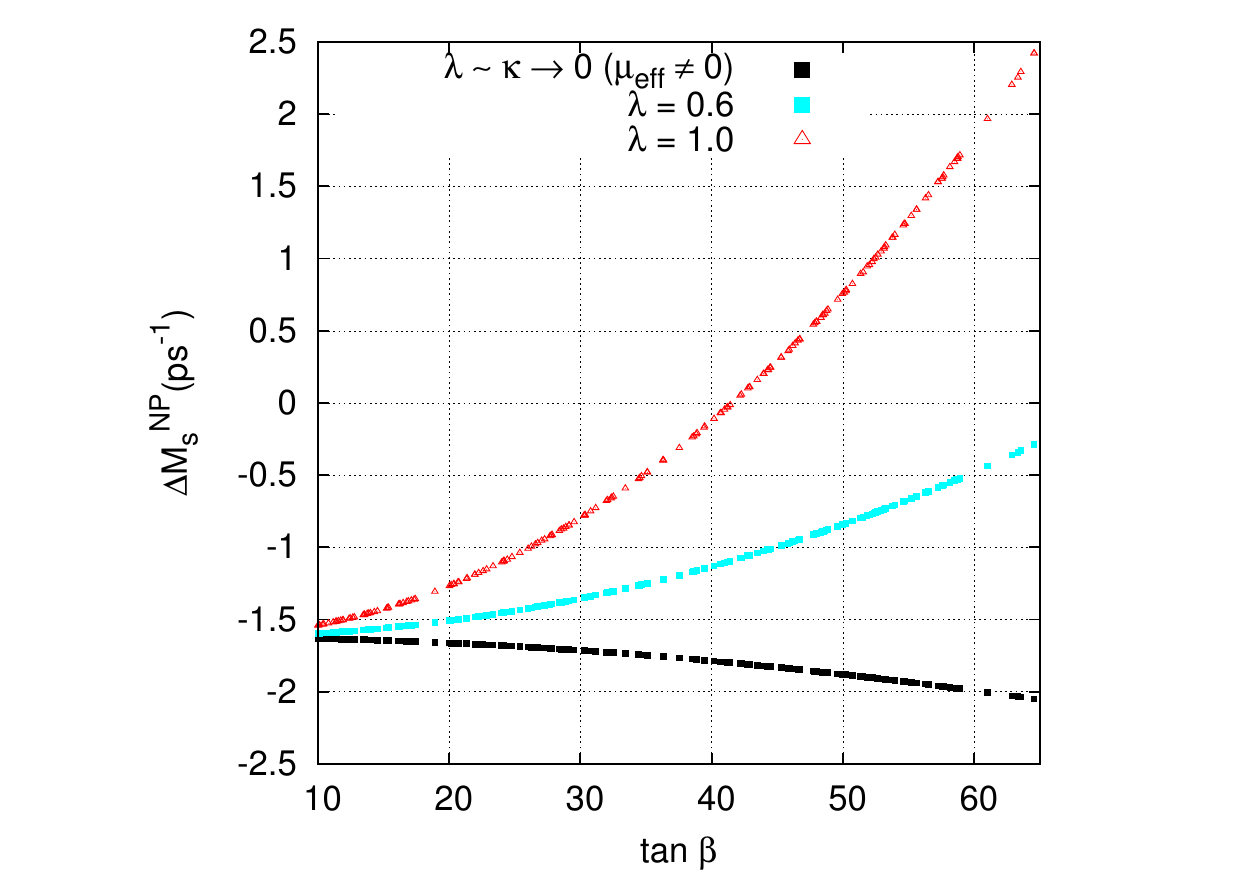}\hspace{-2.7cm}
\includegraphics[height=7.3cm,width=11.1cm,trim={0.cm 0.37cm 0.cm 0.3cm},clip]{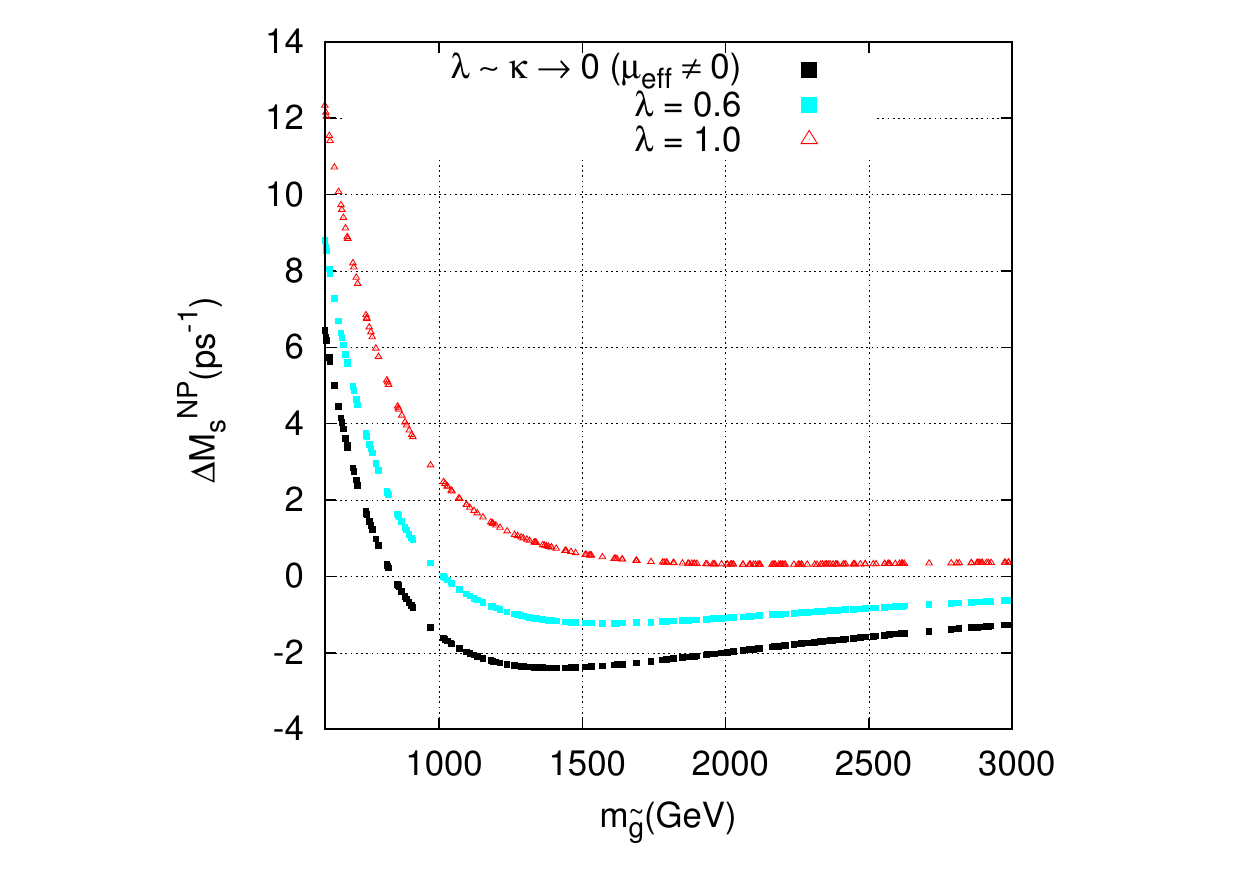}\vspace{0.5cm}
\caption{\small Genuine-NMSSM effects in $\Delta M_s$, understood as deviations with respect to the MSSM predictions under $\tan\beta$ (left) and gluino mass (right), scaling. Input parameters primarily controlling the effect read $(m_D^2)_{ii}=650~GeV, M_S=3 ~TeV, \delta_{RR}^{23}=0.6 $, while $m_{\tilde{g}}=1.1~TeV$ and $\tan\beta =60$ were used for left and right plot, respectively. Cyan line ($\kappa=0.4$) corresponds to perturbative NMSSM up to GUT-scale. Red line ($\kappa=1$) requires UV-completion before GUT-scale, as in $\lambda$-susy models. The black line is the MSSM-limit of the NMSSM model. For other parameters see text. Calculations are performed in mass basis taking into account all contributions.}\label{fig:dms}
\end{figure}\normalsize

In the case of $\Delta M_s$, shown in Fig.\ref{fig:dms}, we consider down-squark flavour violation originating only from $\delta_{RR}^{23}$. This is because, in general, such mass insertions are not strongly constrained by other observables and in particular from $\mathbf{\mathcal{B}}(B\to X_s\gamma)$ which typically sets very strong bounds in $b-s$ mixing phenomena. In left-figure, we vary $\tan\beta$ for a fixed $m_{\tilde{g}}=1.1~TeV$. We notice that a significant splitting between MSSM and NMSSM is induced, which increases with respect to $\tan\beta$. The size of the splitting also increases together with $\lambda\sim \kappa$ parameters. Both behaviours are expected from our previous qualitative discussion on genuine NMSSM-contributions, where the former was associated with the strength of the Yukawa coupling $Y_b$ and the latter with the size of higgsino-singlino mixing (for a fixed $\mu_{eff}=180\, GeV$). The operators responsible for leading genuine-NMSSM contributions in the case of $\delta_{RR}^{23}=0.6$ are found to be only $Q^{SRR}_1, Q^{SRR}_2$, as suggested by Table \ref{tab:mi}, even though for such large values of insertions, the validity of MIA is typically under question\footnote{Under mass insertion scaling, higher order terms in the MIA become increasingly important.}. All other contributions from charged-Higgs, chargino and neutralino-neutralino boxes are found to be negligible.   

In Fig.\ref{fig:dms}-right we examine the stability of the parameter space under gluino mass-scaling in the NMSSM-enhanced $\tan\beta=60$ region. The soft squark mass parameters $\delta_{RR}^{23}, (m_D^2)_{ii}$ (which are effective in the $RR$-induced squark flavour violation scenario) are chosen so that in the experimentally favoured $m_{\tilde{g}}\gtrsim 1.1~TeV$ region, the overall MSSM NP-contribution stays roughly at the $\sim(-10 \%)$ level, as compared to the SM prediction $\Delta M_s^{SM}=19.6 ~ps^{-1}$. We note that significant deviations between MSSM and NMSSM persist even at $m_{\tilde{g}}=3~TeV$ due to the different decoupling behaviour of gluino-gluino and neutralino-gluino contributions. 

\begin{figure}[t]
\hspace{-1.5cm}\includegraphics[height=7.55cm,width=10.6cm]{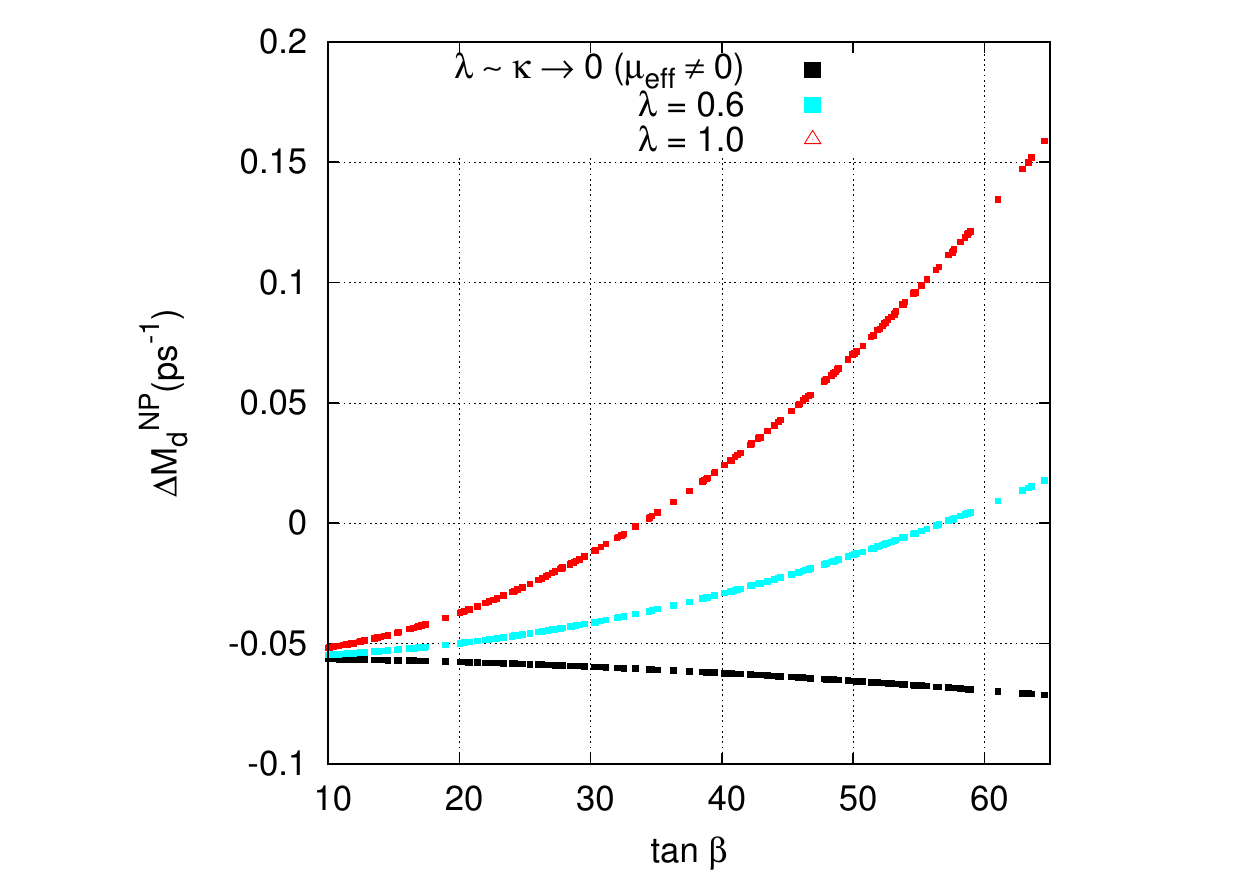}\hspace{-2.7cm}
\includegraphics[height=7.3cm,width=11.1cm,trim={0.cm 0.37cm 0.cm 0.3cm},clip]{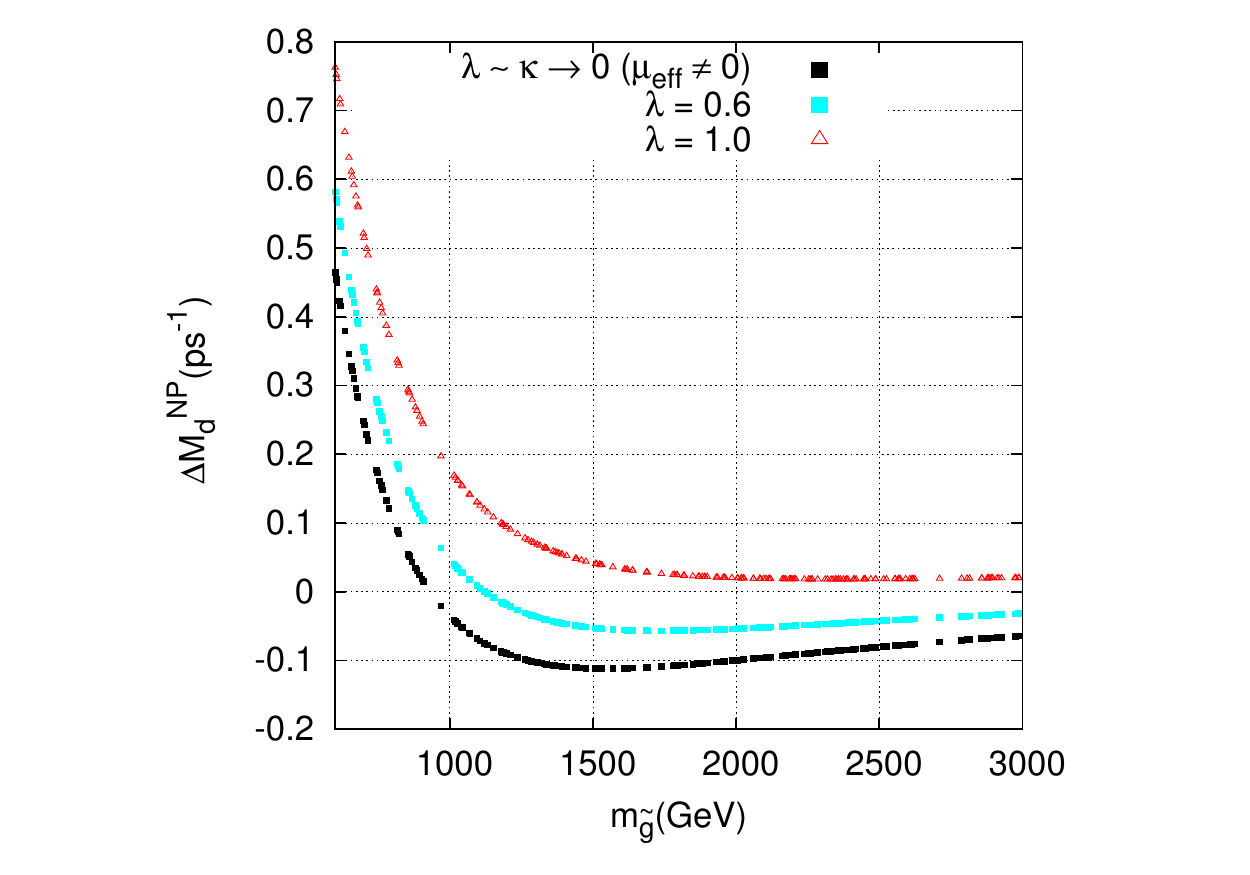}\vspace{0.5cm}
\caption{\small Genuine NMSSM-effects in $\Delta{M_d}$ with all input parameters as in Fig.\ref{fig:dms} besides down-squark flavour violation which is now induced  through $\delta_{RR}^{13}=0.2$. }\label{fig:dmd}
\end{figure}\normalsize

The genuine-NMSSM effect is more consistently understood as the relative shift between MSSM and NMSSM predictions with respect to the SM value. We therefore introduce a \emph{deviation} measure, defined for $B_q$ mixing as 
\begin{eqnarray}
\delta (\Delta M_q)_{N-M}\equiv{ (\Delta M_q^{NP})_{\textnormal{\tiny NMSSM}}-(\Delta M_q^{NP})_{\textnormal{\tiny MSSM}}\over\Delta M_q^{SM}},\label{eq:nmssm-div}
\end{eqnarray} 
since in this quantity accidental cancellations, which are known to commonly occur for gluino-gluino contributions \cite{Crivellin:2010ys,Arhrib:2001gr} or other, become irrelevant. Applying this to the case where the $\delta_{RR}^{q3}$ is the only source of squark flavour violation, one can also check that $(\delta^{q3}_{RR})^2$ factors out from both 
terms in the numerator of eq.\eqref{eq:nmssm-div} in squark MIA.
Thus, squark insertions besides controlling the overall magnitude of NP-contributions, they also control the size of the deviation $\delta (\Delta M_q)_{N-M}$.

In the case of $\Delta M_d$ shown in Fig.\ref{fig:dmd} the situation is analogous and the previous discussion applies here, as well. Both figures display the same qualitative behaviour as before, however one now has to consider $\delta_{RR}^{23}=0.2$ in order to achieve a NP-contribution for MSSM of order $(-20\%)$, as compared to $\Delta_{M_d}^{SM}=0.63 ~ps^{-1}$. In this case, where a larger NP-contribution has been considered, the genuine-NMSSM measure takes higher values, resulting in larger splittings in case of $\tan\beta$-scaling (Fig.\ref{fig:dmd}-left) and larger shifts in case of gluino mass-scaling (Fig.\ref{fig:dmd}-right). This is expected from Eq.\eqref{eq:nmssm-div} and the fact that $(\delta_{RR}^{q3})^2/\Delta M_q^{SM}$ is taken larger in the $q=d$ case. If instead $\delta_{RR}^{13}\approx 0.2/\sqrt{2}$ was considered, corresponding to a  $\sim(-10\%)$ MSSM background as in $\Delta M_s$ case,  the size of the deviation would be practically the same as before.

\begin{figure}[t]
\begin{center}
\includegraphics[height=6.25cm,width=9.5cm]{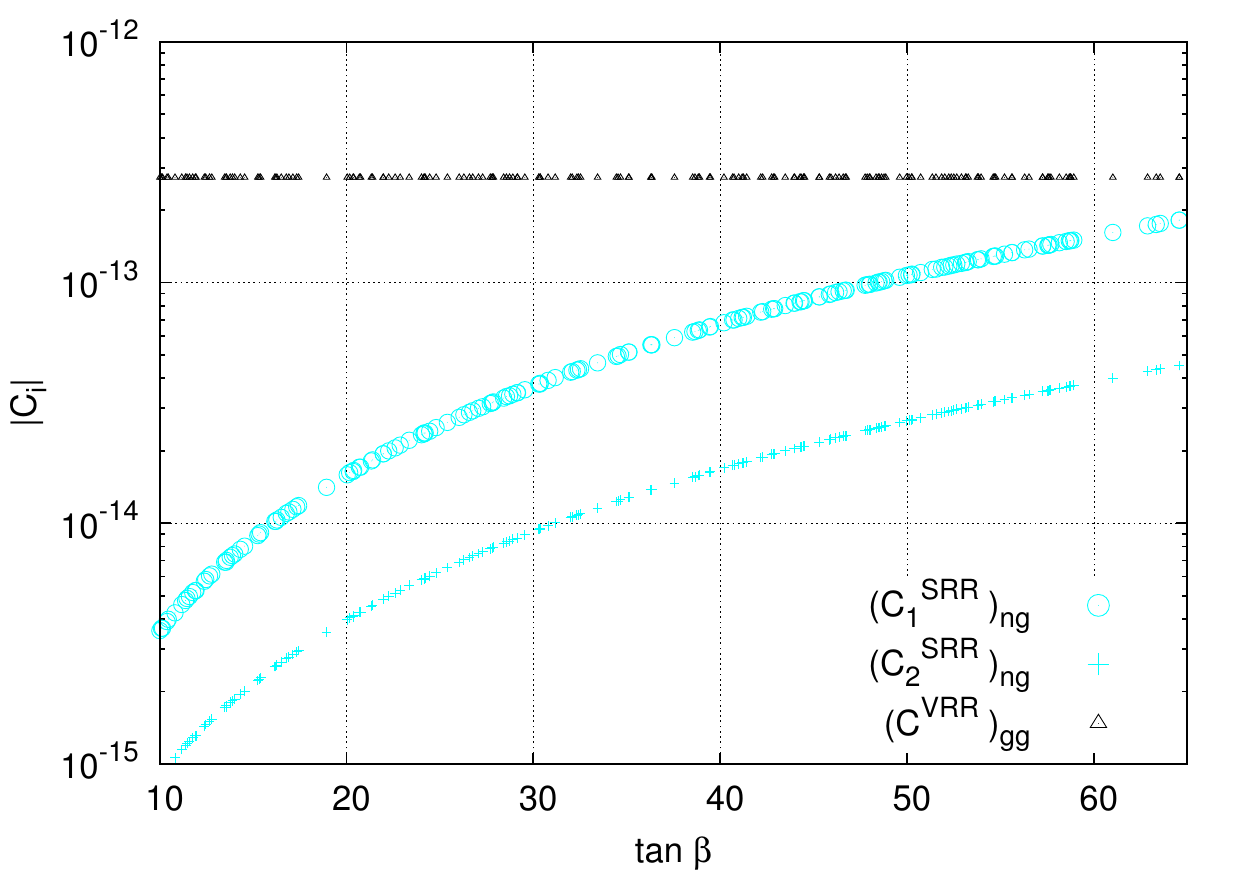}
\caption{\small Leading pure-MSSM $(C^{VRR})_{gg}$ and genuine-NMSSM $(C_1^{SRR})_{ng},(C_2^{SRR})_{ng}$ Wilson Coefficients at the matching scale, for $\lambda=0.6$ and all other inputs as in Fig.\ref{fig:dmd}-left. For the non-perturbative case ($\lambda=1$), which is not shown here, the effect is even larger and 
$(C_1^{SRR})_{ng}$ overtakes. }\label{fig:gg-ng}
\end{center}
\end{figure}\normalsize

Having discussed genuine-NMSSM contributions in $\Delta M_s$ and $\Delta M_d$, we now proceed to an analysis of the leading Wilson coefficients in MSSM and NMSSM, at the matching scale, which are essentially the sources for the observed deviations. We focus on the $\Delta M_d$ case but as argued many times in the text, an analysis for $\Delta M_s$ is in straightforward analogy. In Fig.\ref{fig:gg-ng}, we show the behavior of the leading gluino-gluino pure-MSSM  contribution, $|(C^{VRR})_{gg}|$ versus the leading neutralino-gluino genuine-NMSSM contributions $|(C_{1}^{SRR})_{ng}|,|(C_{2}^{SRR})_{ng}|$, under $\tan\beta$ scaling. All inputs have been taken from Fig.\ref{fig:dmd}-left, for $\lambda=0.6$, essentially giving the picture of the same effect in the language of WC but at the matching scale. The QCD-running and the relevant bag-factors bring some non-negligible effects in the observables, nevertheless they cannot modify the qualitative characteristics we discuss here. The first thing to notice in Fig.\ref{fig:gg-ng} is the domination of the pure-MSSM $|(C^{VRR})_{gg}|$ at low $\tan\beta$. As already mentioned, in the low $\tan\beta$ regime the genuine-NMSSM effects are negligible due to small values of $Y_b$.  However, as $\tan\beta$ grows they become significantly enhanced due to their $Y_b^2$ dependence. At the same time the gluino-gluino contribution, being insensitive to $\tan\beta$, remains constant. This is also seen in Fig.\ref{fig:dmd}-left, where a minor deviation of MSSM with respect to $\tan\beta$, originates only from the other pure-MSSM neutralino-gluino contributions.

A final general remark concerns other possible flavour violating sources in the down-squark sector. In our numerical study we have focused on genuine-NMSSM effects associated only with $\delta_{RR}^{q3}$, for  $\Delta M_{q}$. This approach was motivated by the fact that such insertions are known to be less sensitive to other flavour observables. In this sense, they were not expected to severely constrain our parameter space, a fact which we have also confirmed numerically. This allowed for a common study and a comparison of the effect in $B_s,B_d$ -mixing. However one can always use Table \ref{tab:mi} as a guide to other flavour violation scenaria. In particular, we have checked that a same order effect can in principle arise from the $LL$-sector, namely for $\delta_{RR}^{q3}$ replaced by $\delta_{LL}^{q3}$, since the mechanism for generating genuine-NMSSM contributions is essentially the same (\emph{i.e.,} crossed box contributions proportional to $g_3^2 Y_b^2$, as in Fig.\ref{fig:nmssm}). Nevertheless, such large insertions are strongly constrained by $\mathbf{\mathcal{B}}(B\to X_q\gamma)$ and  $\mathbf{\mathcal{B}}(B\to \mu^+ \mu^-)$ and  typically larger pure-MSSM effects are present in the neutralino-gluino contributions. In the case of $(\delta_{LR(RL)}^{q3})$ a more model-independent argument is effective, since due to $\tan\beta$ suppression in the relevant mass entries ($\sim v_d A_D$), large mass-insertions cannot be easily reached without violating the relevant bounds arising from Charge and Color Breaking minima.

\subsection{NMSSM contributions in Double Penguins}
\subsubsection{General framework for Double Penguin effects in NMSSM}\label{generaldp}
\begin{figure}[t]
\centering
  \includegraphics[trim={3.3cm 22cm 0 3.8cm},clip]{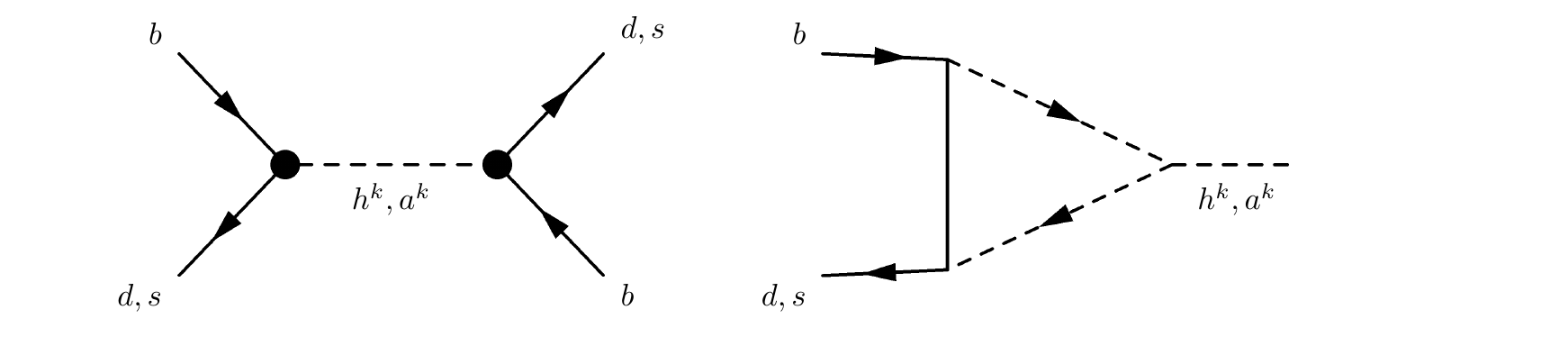}
  \caption{\small Double penguin diagrams (formally two-loop) on the left, induced by one-loop effective Yukawa couplings as the one shown on the right, scaling as $\sim(\tan\beta)^4$ and thus potentially significant for $\Delta F=2$ observables in both MSSM and NMSSM models.}
\label{fig:dp}
\end{figure}
\normalsize

As has been long noticed for MSSM at large $\tan\beta$, certain two-loop diagrams, commonly referred to as \emph{Double Penguins} (DP) \cite{Buras:2003td,Huang:1998bf,Buras:2002vd}, can dominate over  box contributions and even send $\Delta F=2$ observables far beyond experimental bounds. Such diagrams, as shown in Fig.\ref{fig:dp}, involve the exchange of the CP-even and CP-odd scalars and despite being two-loop suppressed, they give significant contributions due to their \emph{chiral-enhancement} through positive powers of $\tan\beta$. Even in scenaria without any genuine source of squark flavor violation, such effects can become important \cite{Buras:2003td,Buras:2002vd,Buras:2001mb}. There is already an extensive literature covering this subject in MSSM\cite{Dedes:2002er,Buras:2002vd,Buras:2002wq,Dedes:2003kp} as well as several focused studies for NMSSM and its variations\cite{Hodgkinson:2008qk,Crivellin:2015dta,Hiller:2004ii,Domingo:2007dx}. Thus, in this section, our discussion stays mainly at the qualitative level. Once again, we isolate the origin of genuine-NMSSM contributions in DP effects, in order to discuss possible scenaria where these are expected to be significant.

Our analysis follows closely the \emph{decoupling-limit} method \cite{Crivellin:2015dta,Crivellin:2011jt} applied here to the case of $Z_3$-invariant NMSSM and taking into account the stringent bounds on Higgs masses and NMSSM parameters, set by the  minimization conditions and phenomenological considerations in the large $\tan\beta$ regime. As has been noted in \cite{Crivellin:2015dta} the NMSSM effects, related to the presence of extra singlet scalar states in the Higgs sector, are \emph{typically} subleading. Essentially, as also discussed here, they are suppressed by at least one power of $v_d/v_s$ with respect to the leading $\tan\beta$-enhanced contributions mediated by the heavy Higgs doublets. Nevertheless, under certain conditions this suppression can be compensated, as discussed in the following section.   

It is useful to display the formulas for the WC in B-meson mixing, induced by DP-contributions in NMSSM. In the operator basis of \eqref{bbops}, they read (\emph{no sum over $i$}),
\begin{align}
&C_1^{SLL}= -{1\over 4} \sum_{k=1}^3\Bigg( {Y^{{i3k}}_{h}Y^{{i3k}}_{h} \over m_{h}^k m_{h}^k}-{Y^{{i3k}}_{a} Y^{{i3k}}_{a}\over m_{a}^k m_{a}^k} \Bigg)^*\\
&C_1^{SRR}= -{1\over 4} \sum_{k=1}^3\Bigg( {Y^{{3ik}}_{h}Y^{{3ik}}_{h} \over m_{h}^k m_{h}^k}-{Y^{{3ik}}_{a} Y^{{3ik}}_{a}\over m_{a}^k m_{a}^k} \Bigg)\\
&C^{SLR}= -{1\over 2} \sum_{k=1}^3\Bigg( {(Y^{{i3k}}_{h})^* (Y^{{3ik}}_{h}) \over m_{h}^k m_{h}^k}+{(Y^{{i3k}}_{a})^* (Y^{{3ik}}_{a})\over m_{a}^k m_{a}^k} \Bigg)
\end{align}     
where $m_h^k,m_a^k$ are the scalar and pseudoscalar masses of the Higgs fields $h^k,a^k$ (in mass basis), respectively. The effective flavour violating Yukawa couplings are defined through the Lagrangian terms, 
\begin{equation}
{1\over \sqrt{2}} Y_h^{3ik }(\bar b_L  q^i_R) \, h^k + {i\over \sqrt{2}} Y_a^{3ik }(\bar b_L  q^i_R) \, a^k + h.c.
\end{equation} 
and as before $i=d,s$ for $B_d,B_s$ mixing, respectively. Following the decoupling limit method of \cite{Crivellin:2015dta} which allows to isolate and resum the chirally enhanced contributions in NMSSM, one can parameterize the WC in the more convenient form,
\begin{align}
&C_1^{SLL}= -{ (\epsilon^{i3}_d)^* (\epsilon^{i3}_d)^*\over 4}\Big({v_u^2\over 2}\Big) \bar\delta_F \label{C1SLLDP}\\
&C_1^{SRR}=-{(\epsilon^{3i}_d)(\epsilon^{3i}_d)\over 4} \Big({v_u^2\over 2}\Big) \delta_F\label{C1SRRDP}\\
&C^{SLR}= -{(\epsilon^{i3}_d)^*(\epsilon^{3i}_d)\over 2} \Big({v_u^2\over 2}\Big) s_F\, ,\label{CSLRDP}
\end{align} 
where we have defined, 
\begin{align}
\bar\delta_F\equiv \sum_{k=1}^3\Big( \bar F_h^k \bar F_h^k -\bar F_a^k \bar F_a^k\Big)\label{fhk2bar}\\
 \delta_F\equiv \sum_{k=1}^3\Big( F_h^k F_h^k -F_a^k F_a^k\Big)\label{fhk2}\\ 
 s_F \equiv\sum_{k=1}^3\Big( \bar F_h^k F_h^k +\bar F_a^k F_a^k\Big)\label{fak2}
\end{align}   
For \emph{real} $\mu_{eff},\lambda$ parameters and $v_s\equiv \sqrt{2}~ (\mu_{eff}/ \lambda )$ the expressions for $F^k_{h(a)},\bar F^k_{h(a)}$ read,
\begin{align}
&\bar F_h^k = {1\over m_h^k} \Big({Z_h^{1k}\over v_d}-{Z_h^{2k}\over v_u} -{Z_h^{3k}(1-\bar x)\over v_s}\Big)\label{fhkbar}\\
&F_h^k = {1\over m_h^k} \Big({Z_h^{1k}\over v_d}-{Z_h^{2k}\over v_u} -{Z_h^{3k}(1-x)\over v_s}\Big)\label{fhk}\\
&\bar F_a^k=F_a^k = {1\over m_a^k} \Big({Z_a^{1k}\over v_d}+{Z_a^{2k}\over v_u} +{Z_a^{3k}\over v_s}\Big)\label{fak}
\end{align}
while if one allows for CP-violation phases, the expressions must be modified accordingly, as will be discussed shortly. Certain important remarks on the above parameterization are in order.

The fixed indices $1$-$3$ of the rotation matrices $Z_{h(a)}$ refer to the initial CP-even and CP-odd ``flavour" (gauge) bases $(H_d,H_u,S)$ and  $(A_d,A_u,A_s)$. They are defined in our conventions through the relations
\begin{align}
\bm{M^2_{H(A)}}=\bm{Z_{h(a)}~ m^2_{h(a)}~  Z_{h(a)}^\top},
\end{align}
with the explicit form of $M^2_{H(A)}$ given in  \ref{app:hmin} and references therein.
Notice that we keep the initial eigenbasis for the CP-odd sector as well, where the effect of the Goldstone mode is not rotated away. For the purpose of our discussion this is more convenient since it makes clear the correspondence between the initial CP-even and CP-odd eigenstates. One can always use the $R(\beta)$ rotation of \ref{app:hmin} and redefine all relevant Yukawa couplings accordingly, which results to a Goldstone state with vanishing quark flavour-violating couplings. In any case, for large $\tan\beta$,  $R(\beta)\approx I$ and thus to a good approximation $A_u\approx G^0 $, as  discussed in our appendix.   

The $\epsilon_d$ parameters are associated with the elements of the $(\Sigma_d)_{LR(RL)}$ part of the down-quark self energies  and are found to be independent of the genuine NMSSM parameters $\lambda, \kappa , A_\lambda ,A_\kappa$ in the ``decoupling limit". In fact they are \emph{common} in MSSM and NMSSM for $\mu=\mu_{eff}$ and include the chirally enhanced part ($\propto v_u$) of the flavour violation effects. They are well studied and their explicit form can always be taken from\footnote{Although expressed analytically under different conventions, we note that $v_u \epsilon_d^{3i} /\sqrt{2}= \bar v_u\bar\epsilon_d^{3i}$ numerically holds for the same physical model. With this observation one can understand the transformations required for setting all relevant expressions to our conventions. Barred notation refers to the conventions used in \cite{Crivellin:2015dta}.}\cite{Crivellin:2015dta}. Nevertheless, this is not vital for our discussion.

The parameters $\bar x,x$, appearing in eqs.\eqref{fhkbar},\eqref{fhk} apply only to the case of \emph{real} $\mu_{eff},\lambda$. They are associated with additional chirally enhanced effects of the CP-even singlet state $S$ through neutral and charged Higgsino propagator contributions in $\epsilon_d$\cite{Crivellin:2015dta} and act as suppression factors through $(1-\bar x),(1- x)$ in $\bar F^k_{h},F^k_{h}$, respectively. In scenaria of enhanced genuine-NMSSM contributions that we are interested in, they are expected to give subleading effects. 
One can easily understand the typical range for these parameters from their formal definition, obtained by the following procedure:
\begin{itemize}
\item
One first decomposes $\epsilon_d^{3i}$ into higgsino and non-higgsino parts, as 
\begin{align}
\epsilon_d^{3i}= (\epsilon_d^{\tilde{H}^\pm,\tilde H^0})^{3i}+
(\epsilon_d^{\cancel{\tilde{H}^\pm,\tilde H^0}})^{3i}.
\end{align}
In practice every term in the explicit expressions of $\epsilon_d^{3i}$ \cite{Crivellin:2015dta} which depends on $\mu^2$ through the arguments of the $C_0$ loop-function is a (neutral or charged) higgsino term. For the CP-even singlet contributions only, there are extra higgsino related terms associated with $\partial(\epsilon_d^{\tilde{H}^\pm,\tilde H^0})^{3i}/ \partial \mu^2$ (thus independent of genuine-NMSSM parameters, as well) and which are trivially obtained with the replacement rule       
\begin{align}
(\hat \epsilon_d^{\tilde{H}^\pm,\tilde H^0})^{3i}=(\epsilon_d^{\tilde{H}^\pm,\tilde H^0})^{3i}\Big[C_0(\mu^2,...)\to 2\mu^2 D_0(\mu^2,\mu^2,...) \Big]\,.
\end{align}
Then, one can consistently define the dimensionless parameter $x$ through the relation,
\begin{align}
1-x \equiv {\epsilon_d^{3i}+(\hat \epsilon_d^{\tilde{H}^\pm,\tilde H^0})^{3i}\over \epsilon_d^{3i} }=1+ {\Big((\hat \epsilon_d^{\tilde{H}^\pm,\tilde H^0})^{3i}/   (\epsilon_d^{\tilde{H}^\pm,\tilde H^0})^{3i}\Big)\over 1 + \Big((\epsilon_d^{\cancel{\tilde{H}^\pm,\tilde H^0}})^{3i}/ (\epsilon_d^{\tilde{H}^\pm,\tilde H^0})^{3i}\Big) }\, .\label{eq:epsdhat}
\end{align}
The definition of $\bar x$ which is instead related to $(\epsilon_d^{i3})^*$ proceeds in a straightforward manner.
The above procedure can be easily modified to include complex $\mu_{eff},\lambda$ as well, by taking the general expressions from \cite{Crivellin:2015dta} and applying an analogous treatment. This will eventually give rise to a similar effect in the CP-odd singlet sector, as well, with $\bar F^k_{a}\neq F^k_{a}$ and scalar-pseudoscalar mixing in the general CP-violation case\cite{Ellwanger:2009dp,Cheung:2010ba,Graf:2012hh}.
\end{itemize}

By direct inspection on the r.h.s. of eq.\eqref{eq:epsdhat} the parameter $x$ (and $\bar x$) is typically expected positive and small (or zero). The sign can be understood from the sign flip between $C_0$ and $D_0$ loop-functions. The size is understood by the fact that the denominator becomes enhanced for large non-higgsino contributions while the numerator ($\sim |2\mu^2 D_0/C_0|\leq 2$)  takes values close to zero when $\mu$ is much lighter than the other mass arguments in the loop-function. When $x$ and $\bar x$ are dropped, they result in a major  simplification of all relevant expressions with $\bar F^k_{h}= F^k_{h}$ and thus $\bar\delta_F=\delta_F$, holding.

The parameterization of the WC through  eqs.\eqref{C1SLLDP}-
\eqref{CSLRDP} is useful when searching for deviations between MSSM and NMSSM induced through DP-effects.  One starts from a specific NMSSM scenario with a realistic mass spectrum and takes its MSSM limit in the standard fashion ($\lambda\to 0,\, \lambda / \kappa=$fixed). The two models then display different numerical values for $\delta_F,\bar \delta_F,s_F$, reflecting the different Higgs  sectors (mass spectrum and rotation matrices). Since the $\epsilon_d$'s act as common factors, only the aforementioned numerical values control the WC ratios, 
\begin{align}
r^C_i = C_i^{NMSSM}/C_{i}^{MSSM}
\end{align}
and therefore can be used to estimate the deviation in $\Delta F=2$ predictions. Obviously, a deviation in the observables takes into account other effects as well, like RGE running and mixing with other WC. Nevertheless, when the NMSSM parameters $\delta_F,\bar \delta_F,s_F$, dominate over the MSSM ones, then $r^C_i\gg 1$ holds and significant deviations (induced by large genuine-NMSSM effects) are expected\footnote{Clearly $\epsilon_d$'s are also important since as ``common factors"  they can suppress (or enhance) the DP-effects altogether.}.  

\subsubsection{Enhanced genuine-NMSSM effects in Double Penguins}
The parameterization of eqs.\eqref{C1SLLDP}-\eqref{CSLRDP} along with considerations of the Higgs potential in NMSSM (\ref{app:hmin}), allow also for a \emph{qualitative} approach on enhanced genuine-NMSSM contributions in limiting cases. As before, we take real $\lambda,\mu_{eff}$ parameters and for simplicity we further neglect the singlet CP-even suppression factors, by setting $\bar x=x =0$  and thus $\bar\delta_F=\delta_F$ in our qualitative analysis (only).

 We first notice that the value of $M_A$ for $Z_3$-NMSSM at large $\tan\beta$ is severely constrained by the minimization conditions. In fact for large values of $\lambda$ the natural scale of $M_A$ is at the multi-TeV range ($\sim\mu\tan\beta$) while for smaller values all our numerical scans with \texttt{NMSSMTools} suggest that $M_A$ cannot in practice lie far below $\sim 1 \, TeV$. We remind that very small values of $\lambda$ suggest an MSSM-limit model, where all NMSSM effects are expected to decouple. For NMSSM (and MSSM) at large $\tan\beta , M_A$, the heavy doublets\footnote{We loosely refer to mass eigenstates as ``doublets" and ``singlets" primarily for convenience,  using as a label the leading flavour (gauge) state  in the field composition of the respective mass eigenstate. However, since the mixing in NMSSM at large $\tan\beta$ is much smaller than maximal, such a definition is also accurate to a good approximation.} display a strong degeneracy which allows one to safely use the approximation 
\begin{equation}
m_h^1  \approx m_a^1  \approx M_A
\end{equation}
and simplify substantially all relevant expressions.  

The genuine-NMSSM effects in Double Penguins are understood as contributions related to the singlets. They appear in the WC of eqs. \eqref{C1SLLDP}-\eqref{CSLRDP} as deviations in the respective values of $\delta_F(\bar\delta_F),s_F$, between an NMSSM scenario and its MSSM-limit. Depending on the NMSSM Higgs sector, the leading deviations arise either directly from the singlet terms  $ Z^{33}_{h(a)}/(m_{h(a)}^3 v_s) \subset F^3_{h(a)}$ or from the heavy doublet-singlet mixing terms $ Z^{13}_{h(a)}/(m_{h(a)}^3 v_d)\subset F^1_{h(a)}$ since heavy-doublets mediate the leading $\tan\beta$-contributions. As genuine-NMSSM terms, they both vanish in the MSSM-limit. When there is a strong  hierarchy between them, a qualitative analysis can take place. Hence, we classify NMSSM models as the \emph{small-mixing} and the \emph{large-mixing} case, while when both contributions are comparable (moderate-mixing) numerical methods for the evaluation of $\delta_F(\bar\delta_F),s_F$ are preferable. 
 
In our analysis we impose a rough lower bound on the absolute magnitude of genuine-NMSSM contributions, in each case with a different expression. In practice, we require that these dominate over the light doublet ones in $F^{k}_{h(a)}$ (eqs.\eqref{fhkbar}-\eqref{fak}) and thus in the WC. This is because light doublet effects are known to be small (\emph{i.e.} $\tan\beta$ suppressed to heavy doublets) and even subleading to the box contributions. Thus genuine-NMSSM contributions of this order are expected to be unobservable. In this way, we also neglect light doublet contributions in our approximate formulas.

\subsubsection{The small mixing case of NMSSM}
 We first examine the case where the rotation matrices can be approximately expressed as $\mathbf{Z_{h(a)}}= \bm{I}+\bm{\mathcal{O}}(\mathcal{\varepsilon})\approx \bm{I}$. The  corrections $\bm{\mathcal{O}}(\mathcal{\varepsilon})$ which we neglect here, determine the effective range of the small-mixing  analysis, and their size will be discussed later on.

The expressions in this approximation, read
\begin{align}
C_1^{SLL(SRR)}\propto ~~\delta_F \simeq \Big({1\over M_A v_d} \Big)^2 \Big( (1-r_h)^2-(1+r_a)^2\Big),\label{dFsmall}\\
C^{SLR}~~~~~~\propto~~ s_F \simeq \Big({1\over M_A v_d} \Big)^2 \Big( (1-r_h)^2+(1+r_a)^2\Big),\label{sFsmall}
\end{align}
where 
\begin{equation}
r_h\equiv {M_A v_d\over m_h^3 v_s}~,~~r_a\equiv {M_A v_d\over m_a^3 v_s} .
\end{equation}

The aforementioned lower bound for genuine-NMSSM effects to be potentially observable is imposed in the small mixing case through the condition,
\begin{equation}
 \max\{( m_h^3 v_s)^{-1}, ( m_a^3 v_s)^{-1} \}\gg ( m_h^2 v_u)^{-1} ,\label{magnbound}
\end{equation}
and the MSSM-limit of NMSSM  is obtained in this case through $v_s\to\infty$ which decouples all singlet effects.  

For genuine NMSSM effects to be significant, mainly two requirements have to be simultaneously satisfied. First, the absolute magnitude of singlet contributions must remain at an observable level. This means  $(m_{h(a)}^3 v_s)^{-1}$ (implicit in $r_{h(a)}$) must at least satisfy eq.\eqref{magnbound}, with light singlet masses and vev enhancing further the effects. Obviously, the $\epsilon_d$'s are also relevant, since as common factors in the WC of NMSSM and MSSM  they can suppress DP-effects altogether if they are small. We assume implicitly that the flavour violation in the theory is large enough to induce non-negligible $\epsilon_d$'s. 

The second requirement is related to the amount of MSSM-screening in a given model. The $r_{h(a)}$-parameters, being \emph{the ratio of singlet over heavy-doublet contributions} in $F^k_{h(a)}$, control  this screening. When at least one of the two satisfies $r_{h(a)}\gtrsim 1$, the genuine-NMSSM effects are expected leading in all WC contributions. However, due to perturbativity  and phenomenological considerations ($\lambda\lesssim \mathcal{O}(1),\,\mu_{eff}\gtrsim 100 \, GeV$), $v_s$ cannot be much smaller than $v_u$. As a result, singlet contributions $(m_{h(a)}^3 v_s)^{-1}$ are typically subleading,  being suppressed by at least one power of $(v_d/v_s)\lesssim \cot\beta$ as compared to the heavy-higgs doublets $(M_A v_d)^{-1}$. Nevertheless, for $ (m_{h(a)}^3/M_A )\lesssim (v_d/v_s)$  the suppression is compensated (\emph{i.e., $r_{h(a)}\gtrsim 1$}) and NMSSM effects overtake.

The expressions for $\delta_F, s_F$ in eqs. \eqref{dFsmall}, \eqref{sFsmall} also allow to study the mechanisms which induce enhanced singlet effects and classify NMSSM models, accordingly.

\paragraph{Singlet squared term domination: ${\max\{r_h,r_a\}\gg 1}$.\newline}
In this case the genuine-NMSSM effects  dominate in $\delta_F,s_F$  and thus in all WC of eqs.\eqref{C1SLLDP}-\eqref{CSLRDP}. To leading order the relevant expressions simplify further, giving 
\begin{align}
\delta_F \simeq \Big({1\over m_h^3 v_s}\Big)^2- \Big({1\over m_a^3 v_s}\Big)^2 \label{dFsmall1}\\
s_F \simeq \Big({1\over m_h^3 v_s}\Big)^2+ \Big({1\over m_a^3 v_s}\Big)^2 \label{sFsmall1}
\end{align}  
As can be easily seen, both expressions are independent of $M_A$. In fact $M_A$, associated with heavy-doublet contributions is only relevant to the magnitude of pure-MSSM effects, which in this case is subleading and thus neglected in the leading order expressions. For very large $M_A$, heavy doublet effects decouple and the bound of \eqref{magnbound} becomes more effective ($ (m_h^2 v_u)^{-1}\gtrsim (M_A v_d)^{-1} $). 

Due to the restrictions on $v_s$ and the requirement that NMSSM effects appear at an observable level, the condition 
$r_{h(a)}\gg 1$ is expected to be satisfied for very light singlet masses. However, the zero momenta approximation we have applied for the singlet propagators is then inaccurate and singlet masses in all relevant expressions have to be replaced through the Breit-Wigner form of the propagators\footnote{Also in the ratios $r_{h(a)}$ discussed here and in $\hat r_{h(a)}$ which is discussed later.}\cite{Domingo:2007dx}. This essentially induces a resonance effect for $m^3_{h(a)}$ close to $M_{B_q}$. Thus, very large values of $r_{h(a)}$ typically imply at least a mild correlation with these resonance effects. 

\paragraph{Singlet-doublet crossed term enhancement: ${\max\{r_h,r_a\}\sim 1}$.\newline}
In this case a significant amount of MSSM screening is expected since NMSSM and MSSM contributions are comparable by assumption. However, there is a certain mechanism that applies to a wide range of models, including MFV realizations and which can enhance further genuine-NMSSM effects making them leading even in cases where $r_{h(a)}\lesssim 1$.

In order to explain this mechanism we first notice that in eq.\eqref{dFsmall}, $\delta_F$ is in practice genuine-NMSSM due to a cancellation of non-$r_{h(a)}$ terms. In fact this is a well-known suppression mechanism for $C^{SLL(SRR)}_1\propto \delta_F$ in MSSM where the leading heavy-doublet contributions cancel due to degeneracy in $\delta_F$, and the remnants being associated with light-doublet effects are subleading to $C^{SLR}\propto s_F$. However, in NMSSM models the remnants are also associated with singlet effects which can be large, giving $\delta_F^{NMSSM}\gg \delta_F^{MSSM}$.

We may focus on the case ${\max\{r_h,r_a\}\lesssim 1}$ where MSSM screening is large and therefore NMSSM-enhancement is more difficult to appear. Obviously, even larger enhancement is expected when the inequality is reversed and singlet squared terms become relevant. The leading contributions now read,
\begin{align}
&\delta_F \simeq -{2\over M_A v_d}  \Big( {1\over m_h^3 v_s} +{1\over m_a^3 v_s }\Big)\\
&s_F \simeq 2\Big({1\over M_A v_d} \Big)^2 
\end{align}
Contrary to singlet squared enhancement, here genuine-NMSSM effects coming only from $\delta_F$ decouple with $(M_A v_d)^{-1}$ and thus $M_A$ has to be relatively light for them to be non-negligible. However, as $M_A$ becomes light the amount of MSSM-screening grows and therefore the largest effect is expected for models balancing between the two requirements.  

Due to $r_{h(a)}\lesssim 1$ the inequality $(\delta_F/s_F)\lesssim 1$ holds, suggesting that genuine-NMSSM effects are expected subleading. However, this is not always the case. To understand this, one needs to consider the explicit expressions for the WC, in this approximation. Neglecting common factors, these read 
\begin{align}
&|C_1^{SLL}| \propto  \Big|{(\epsilon^{i3}_d)^* (\epsilon^{i3}_d)^*} \delta_F\Big|\\
&|C_1^{SRR}|\propto  \Big|{(\epsilon^{3i}_d) (\epsilon^{3i}_d)} \delta_F\Big|\\
&|C^{SLR}|\propto 2|{(\epsilon^{i3}_d)^*(\epsilon^{3i}_d)}s_F|   
\end{align}       
Since we neglect light doublets, only singlet effects are present in $C_1^{SLL(SRR)}$  through $\delta_F$. These will become leading when they dominate over the pure-MSSM one, $C^{SLR}$. One can then define the WC ratios,
\begin{align}
r_L^C \equiv {|C_1^{SLL}|\over |C^{SLR}|}\simeq  \Big|{(\epsilon^{i3}_d)^*\over 2 \epsilon^{3i}_d}\Big|(r_h+r_a) \\
r_R^C \equiv {|C_1^{SRR}|\over |C^{SLR}|}\simeq  \Big|{\epsilon^{3i}_d\over 2 (\epsilon^{i3}_d)^*}\Big|(r_h+r_a) 
\end{align} 
From the above expressions it becomes clear that when the hierarchy between $\epsilon_d$'s is large enough to compensate  the suppression coming from  $r_{h(a)}\lesssim 1$, then the respective genuine-NMSSM Wilson Coefficient takes over. In fact, a strong hierarchy between the $\epsilon_d$'s is the typical case in many models, including MFV in which $(\epsilon^{i3}_d)^*\gg (\epsilon^{3i}_d)$ due to down-type Yukawa couplings associated with different generations. This eventually gives large values for $r^C_L$ and thus genuine-NMSSM domination driven by $C_1^{SLL}$.

\paragraph{Large MSSM-screening: ${\max\{r_h,r_a\}\ll 1}$.\newline}
This is the worst case scenario for NMSSM deviations in $\Delta F=2$ observables. The amount of MSSM screening is so large that genuine NMSSM effects become in practice unobservable. Nevertheless, this is the typical scenario for NMSSM models with singlet masses heavier than  $\sim 100 \,GeV$ or large $v_s$, independent of the flavour violation in the given model.

\subsubsection{The large mixing case and beyond}
What is  small or large mixing essentially depends on the size of the subleading corrections to the previous approximation, $\bm{Z_{h(a)}}= \bm{I} +\bm{\mathcal{O(\varepsilon)}}\approx \bm{I}$. The leading singlet mixing effect is associated with $Z_{h(a)}^{13}/ (v_d m_{h(a)}^3)$. Any other mixing term in eqs.\eqref{fhkbar}-\eqref{fak} is suppressed by at least one power of $\cot\beta$ (or $v_d/v_s$) as compared to this. One can formally define the large mixing case as 
\begin{equation}
\max\{{ Z_h^{13} (  m_h^3 v_d)^{-1} },{ Z_a^{13}( m_a^3 v_d)^{-1}}\}  \gg  \max\{(m_h^3 v_s )^{-1},(m_a^3 v_s)^{-1}\} 
\end{equation}
while for the previous \emph{small mixing} case, the inequality is reversed. 

The relevant expressions in the large mixing case are,
\begin{align}
&C_1^{SLL(SRR)}\propto ~~\delta_F \simeq \Big({1\over M_A  v_d}\Big)^2\Big(({1+\hat r_h})^2-({1+\hat r_a})^2\Big)~~~~\\
&C^{SLR}~~~~~~\propto ~~s_F \simeq \Big({1\over  M_A v_d }\Big)^2\Big(({1+\hat r_h})^2+({1+\hat r_a})^2\Big)~~~~
\end{align}
where now $\hat r_h\equiv {M_A Z_h^{13}\over m_h^3},\hat r_a\equiv {M_A Z_a^{13}\over m_a^3} $. We have also set $Z_{h(a)}^{11}\approx 1$ since even in the large mixing case, the angles are necessarily small for a consistent Higgs NMSSM spectrum at large $\tan\beta$.

The formal condition for NMSSM effects to be potentially  non-negligible, reads here 
\begin{equation}
\max\{{ Z_h^{13} ( m_h^3  v_d)^{-1} },{ Z_a^{13}( m_a^3  v_d)^{-1}}\}  \gg ( m_h^2 v_u)^{-1}
\end{equation}
and the MSSM-limit of NMSSM appears through $Z_{h(a)}^{13}\to 0$. 

Besides signs, which are determined by the sign of $Z_{h(a)}^{13}$ and the different definitions of $\hat r_{h(a)}$ everything else is identical to the small mixing case. Even an analogous dependence on $v_s$ is present in the ratios $\hat r_{h(a)}$, however here implicit in the rotation matrix elements $Z_{h(a)}^{13}$.  Clearly, the range for the rotation matrix element satisfies $Z_{h(a)}^{13}\gg (v_d/v_s)$ for our approximation to be effective. However, typically, it cannot exceed $\cot\beta$ by far in models with a consistent NMSSM Higgs sector. Under this observation one can also perform here a qualitative discussion, in a straightforward analogy to the small mixing case, arriving essentially at the same conclusions.

\begin{figure}[t]
\begin{center}
\includegraphics[height=8.25cm,width=11.5cm]{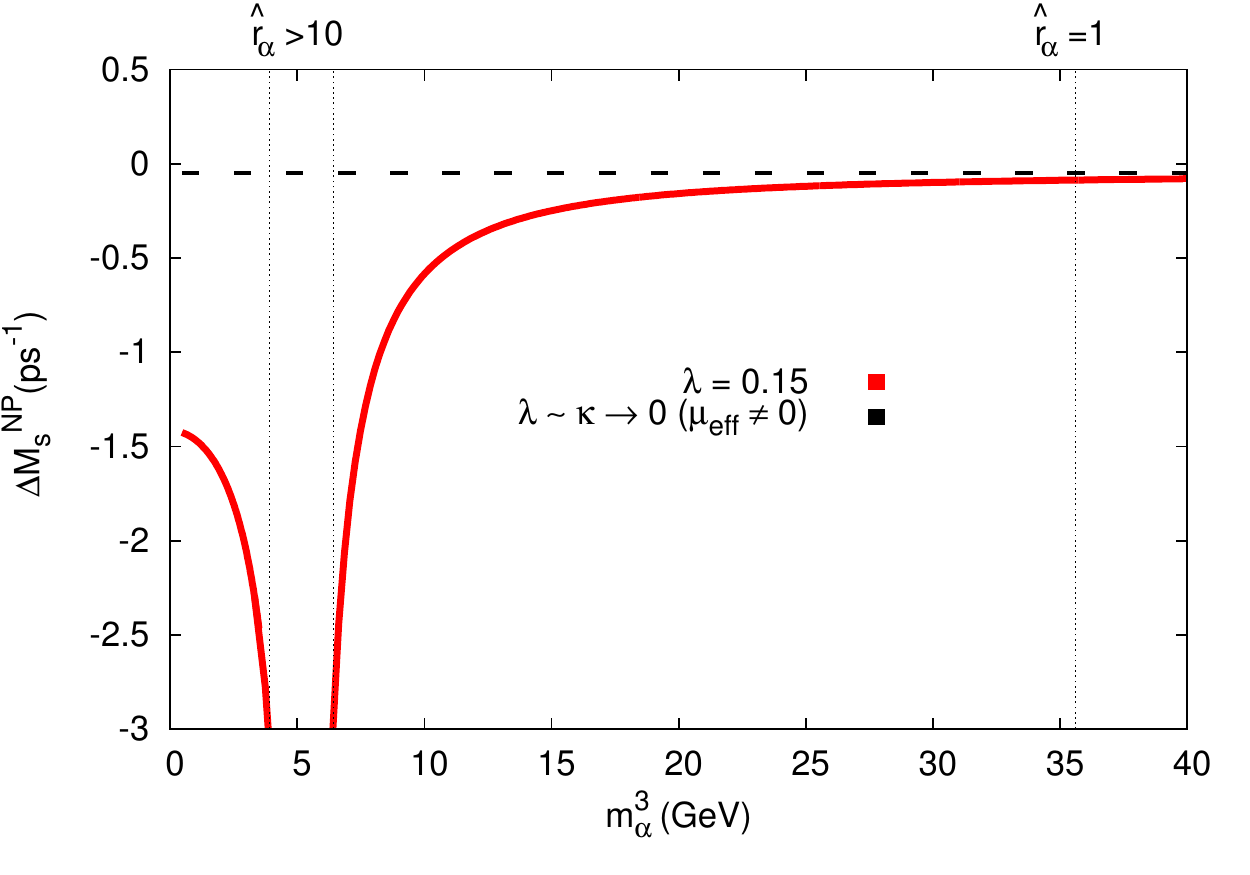}
\caption{\small MSSM (dashed) and NMSSM (red) contributions in $\Delta M_s^{NP}$, under CP-odd mass scaling of the singlet-like eigenstate and driven by $|C^{SLL}_1|\gg |C^{SLR}| $ in the enhancement region.  As the singlet CP-odd mass $m_a^3$ closes to the resonance ($M_{B_s}$), then the squared singlet CP-odd contributions dominate ($\hat r_{a}\gg 1$) and the size of the effect increases rapidly, sending $\Delta M_s^{NP}$ far beyond experimental bounds. The CP-even singlet mass, taken here as an output, remains always heavy.}\label{fig:dms-dp}
\end{center}
\end{figure}
\normalsize

In Fig.\ref{fig:dms-dp} we show a typical behaviour of the large mixing case. We note however that the recent measurements in $B_q\to \mu^+ \mu^-$, the constraints from $B\to X_s \gamma$ and the requirement to fit a SM-like Higgs mass 
in $Z_3$-NMSSM models impose severe constraints on the physical parameter space of their MSSM-limits. As a result, the MSSM background in many cases is effectively zero and significant NMSSM effects arise only for large $\hat r_a$. Genuine-NMSSM contributions here appear at an observable level for a singlet CP-odd mass below $\sim 15~GeV$. For this plot we have considered a degenerate soft squark-spectrum at $M_S=2~TeV$ and 
\begin{eqnarray}
&\tan\beta=50, \mu = 120~GeV, A_t=3~TeV,\nonumber\\
 &M_A=1~TeV, M_{1}=M_2=2~TeV, m_{\tilde g}=1.1 TeV,
\end{eqnarray}
with all mass insertions set to zero.

Having covered the two limiting cases of small and large mixing, the moderate mixing case is expected to carry at least similar qualitative characteristics. As discussed, a certain enhancement mechanism must be effective (\emph{i.e.,} resonance, hierarchy in $\epsilon_d$'s) in order for genuine-NMSSM effects to become leading. Since observable effects are associated in one way or another, with a light singlet spectrum and certain additional conditions (\emph{i.e.,} value of $M_A,v_s$) we conclude that in the vast majority of NMSSM models with heavy singlets, genuine NMSSM-effects are either very subleading or negligible. For NMSSM models with light singlets, however,  Double Penguin genuine-NMSSM contributions are potentially observable. Nevertheless, significant deviations are expected when all enhancement requirements of the respective mechanism are satisfied and, obviously, when the parameter space is unconstrained by other observables.


\section{Upper bounds on new physics in $\Delta F=2$ for MFV models at low $\tan \beta$ in MSSM and NMSSM }\label{3}
In this section we focus on the $U(2)^3$ and $U(3)^3$ 
MFV-scenaria in MSSM and NMSSM at low $\tan \beta$. As has become clear from our
previous discussion, at low $\tan \beta$ both models (independent of MFV assumption)
give effectively the same predictions for the $\Delta F=2$
observables, as long as their common susy-parameters lie in the physical parameter space. However, after the recently discovered 125 GeV Higgs by CMS and ATLAS, this is no longer the case. LHC has imposed severe constraints on  MSSM at low $\tan\beta$, allowing it to be realized through hMSSM scenaria.
In NMSSM this situation is substantially more relaxed. 

Here, we follow a different approach on $\Delta F=2$ observables, as compared to our previous analysis. We use the different lower bounds on charged Higgs masses in the two models as a way to distinguish between hMSSM and NMSSM at low $\tan\beta$, irrespective of the different squark scales to which each model is eventually associated. This is understood when noticed that the dominant contributions in MFV scenaria at low $\tan\beta$ originate from charged Higgs box diagrams, to which the squark spectrum is irrelevant. The flavour violation mechanism in these diagrams is already fixed by the CKM-matrix and the up-quark masses. Therefore, what is only relevant is the charged Higgs mass which, for a fixed $\tan\beta$, essentially controls the magnitude of NP-contributions.  
In what follows we systematically take into account the limits from various Heavy-Higgs searches along with the Higgs observables and turn them into (different) constraints on the $ m_{H^{\pm}} -
\tan \beta$ planes of hMSSM and NMSSM. We find that a lighter charged Higgs mass is in general allowed in NMSSM, which eventually translates into a larger upper bound for NP-contributions in $\Delta F=2$ observables, as compared to hMSSM.

In LHC Run-I, CMS and ATLAS have collected more 
than $\rm 20~ fb^{-1}$ data, and looked for scalars in various topologies for mass scales up to 1 TeV. In particular, a search for charged Higgs bosons in the 
channel $ t \rightarrow b H^{+} (H^{+} \rightarrow \tau^+ \nu)$ 
has been performed by both CMS and ATLAS experiments \cite{CMS:2014cdp,Aad:2014kga} 
while the search for heavy neutral scalars  
in the channels $H \rightarrow ZZ ~(ZZ \rightarrow llll,llqq)$ \cite{Aad:2015kna,Khachatryan:2015cwa} and 
$A \rightarrow hZ~ (h \rightarrow b \bar b,\tau \tau, Z\rightarrow
ee,\mu\mu)$ \cite{Aad:2015wra,Khachatryan:2015lba} has reached $m_{H,A}\sim 1~ TeV$. 
Due to the non-observation of any new scalar, upper limits 
have been set at $95 \%$ CL 
on $ \sigma \times BR$ for each mode. 
These limits along with the measurements of 
Higgs observables can be turned into 
constraints on the $m_{H^{\pm}}-\tan \beta$ planes of 
MSSM and NMSSM. 
They are found to act differently on the two models, since the $BR$ patterns for 
$H^{\pm},H,A$ are essentially distinct due to the following reasons: 
\begin{itemize}
\item In NMSSM the role of the heavy scalars $A(H)$ can be played by multiple 
states\footnote{In this section, we refer to scalars $A_i,H_i$ with index $i$ following an increasing-mass order.}, \emph{i.e, $ A_1,A_2$ $(H_2,H_3)$}.
Depending upon the mixing with the singlet (which in turn affects
its couplings to the fermions and the bosons) the BR predictions 
to a given mode can be drastically different from the MSSM.
\item
Due to the presence of extra CP-even and CP-odd states in NMSSM, 
there exist additional Higgs-to-Higgs decays 
of $ H^{\pm}, H, A $ (to $\phi \phi$ and $\phi V$,  where $\phi$ is 
a scalar or pseudoscalar and $V$ represents $W^{\pm}$ or $Z$) , which become effective when kinematically allowed. 
\end{itemize}
\begin{figure}[t]
\centering
  \includegraphics[width=7cm,height=6cm]{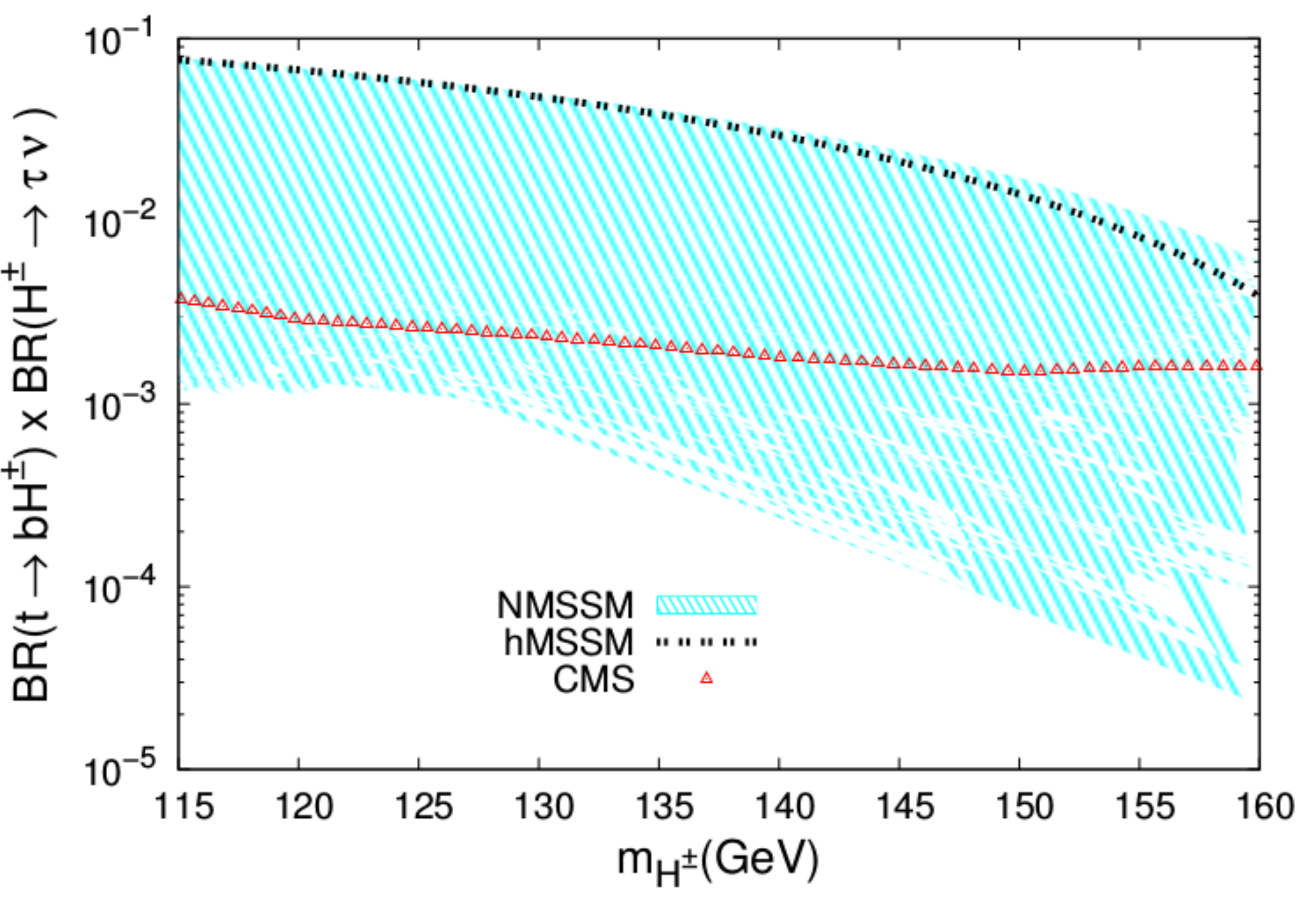}
  \includegraphics[width=7cm,height=6cm]{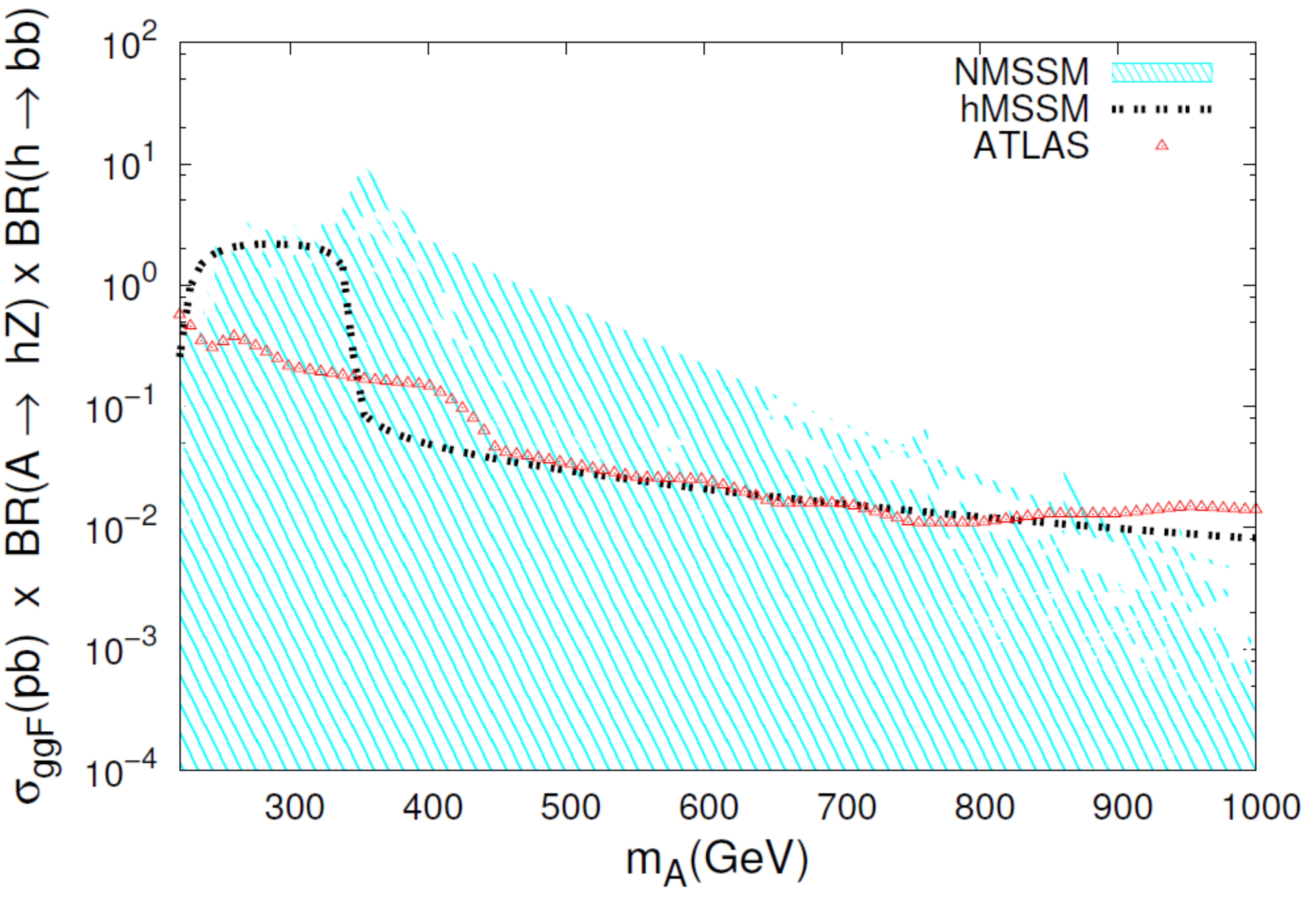}
 \caption{\small The predicted rate for hMSSM is given by the black curve,
cyan represents the NMSSM predictions at $\tan\beta =2$.
and the red line is the experimental upper limit at $95\%~ CL$.}
\label{fig:htanu}
\end{figure}\normalsize

As an illustrative example of the distinct constraints obtained in the two models,
 we compare the various rates 
 at $\tan \beta =2$ where parameters are varied according
to 
\begin{eqnarray}
&\lambda ~:~ 10^{-4}-0.65, ~  \kappa ~:~ 10^{-4} - 0.65, \nonumber\\ 
&\mu ~:~ 0.2-2~TeV,  \ \ A_{\kappa} ~:~ (-0.2) - 2~TeV, \nonumber\\  
 &m_{Q_3} = m_{U_3} = m_{D_3} ~:~ 1~TeV, \nonumber \\
  &  M_2 =2~M_1= 0.5~TeV,  \ \ M_3 = 1.5~TeV.
\label{nmssmpara}
\end{eqnarray}
 Note that for 
the CP-odd Higgs, $A$, only the dominant production in gluon gluon fusion (ggF)
is assumed and we take inclusive production cross section for $H$. The 
branching ratios involved are calculated using 
{\tt HDECAY} \cite{Djouadi:1997yw} for hMSSM and 
{\tt NMSSMTools} \cite{Das:2011dg} for NMSSM. 
The production cross section for heavy scalars in gluon fusion at $\sqrt{s} = 8~ TeV$ 
is computed using the program {\tt SusHi} \cite{Harlander:2012pb,Bagnaschi:2039911}. 

In Figure \ref{fig:htanu} (left) we present  
$BR(t \rightarrow b H^{+})\times BR(H^+ \rightarrow \tau^+ \nu)$ vs. 
$m_{H^\pm}$ for hMSSM (black line)  and NMSSM (cyan area)
along with the CMS upper limit at $95\%$ CL (red line).
As easily noticed, hMSSM predictions lie far above the experimental 
upper limit. This is because in hMSSM, $H^{+} \rightarrow \tau \nu$ 
is the dominant mode for low $\tan \beta$ and light charged Higgs mass and in addition the $BR(t \rightarrow b H^{+})$ is almost $100 \%$.   
But this is not the case for NMSSM, even though the charged 
Higgs couplings to up and down type fermions are same as in MSSM.
For a given charged Higgs mass, the partial decay widths 
in the two models are equal, satisfying
${\Gamma_{H^{+}\rightarrow \tau^{+}  \nu}^{N}}/{\Gamma_{H^{+}
\rightarrow \tau^{+}  \nu}^{M}}=1$.
Nevertheless, in the presence of much lighter $A_1$ or $H_1$ states, 
the additional $H^{\pm} \rightarrow W^{\pm} H_{1}$, $~W^{\pm} A_1 $ 
modes may overtake  $H^{+}\rightarrow \tau^{+} \nu$, 
when kinematically allowed. As a result, the rate for NMSSM becomes
widely spread around the experimental upper limit.

In Figure \ref{fig:htanu} (right) we present the rate
$  \sigma_{ggF}(A) \times BR(A \rightarrow hZ) \times BR(h \rightarrow b \bar b)$ for hMSSM and  
NMSSM along with the ATLAS upper limit at $95 \%$ CL. 
In hMSSM at low tan $\beta$ there can be a departure 
from the decoupling limit making the coupling $g_{AhZ}$ non-negligible. 
As a result the rate increases together with $BR(A \rightarrow hZ)$, becoming significant for $m_h + m_Z  \lesssim m_A \lesssim 2 m_t$. Beyond the top quark threshold,
the decay mode $A \rightarrow t \bar t$ overtakes. In order to probe this region a search for scalar resonances in $ t \bar t$ final state is required.
For NMSSM, the rate is not always high even for $m_h + m_Z  
\lesssim m_A \lesssim 2 m_t$. As mentioned before this is
due to presence of additional Higgs-to-Higgs decays and also because 
the role of $A(h)$ can 
be played by multiple states $A_1,A_2(H_1,H_2)$.
For example if $A=A_2$ and $h=H_2$ the decay modes $A_2 \rightarrow H_1 Z$ and 
$A_2 \rightarrow H_1 A_1$ can also overtake the $A_2 \rightarrow H_2 Z$ decay mode, 
resulting in a large variation in the rate as indicated by the cyan area. 

\begin{figure}[t]
\centering
  \includegraphics[width=0.5\textwidth]{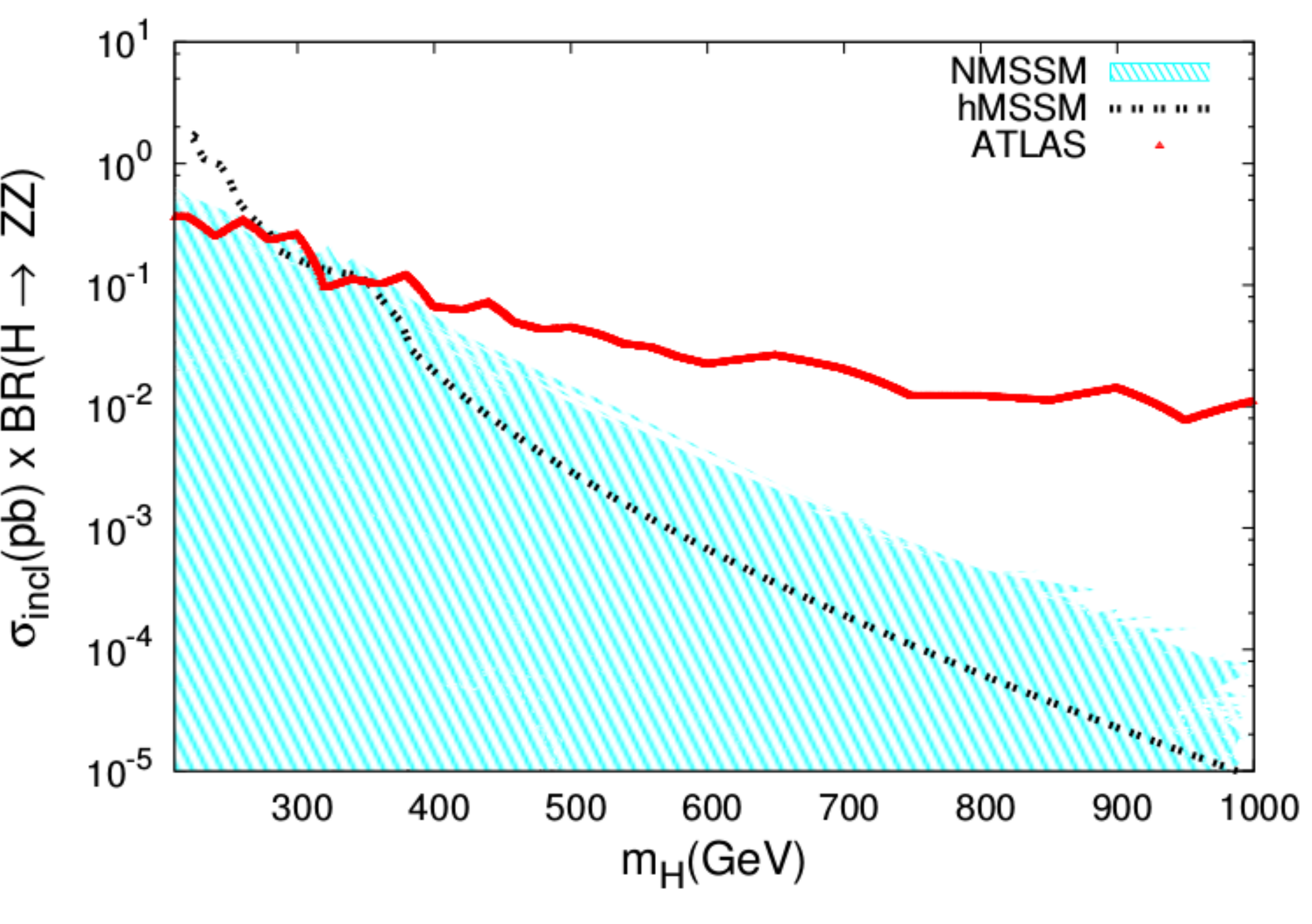}
  \caption{\small Same conventions as in Fig.\ref{fig:htanu}
 }
\label{fig:hzz}
\end{figure}

In Fig. \ref{fig:hzz} we present the rate $\sigma_{\rm inc} 
\times BR(H \rightarrow ZZ)$ 
for hMSSM and NMSSM. As before, the red line 
corresponds to the ATLAS upper limit at $95 \%$ CL. 
We note that for $m_H \lesssim 250$ GeV, 
the predicted rate in MSSM is higher than the experimental limit, 
and therefore the corresponding area is excluded at $95\%~ CL$. 
However due to analogous reasons as before,
only a small portion of NMSSM region is excluded for 
the same mass range.

\normalsize


\subsection{Meson anti-meson mixing and direct search constraints}
Once we take into account the lower bounds on the masses of sparticles
from direct searches, the contributions from gluinos, neutralinos and the gaugino part of charginos, become negligible for MFV-MSSM at low $\tan \beta$. 
The NP contributions then involve dominant charged Higgs   
and subleading chargino diagrams with the latter related only to the (charged) higgsino state\cite{Barbieri:2014tja}. As previously mentioned, the situation for NMSSM in general and MFV-NMSSM in particular, is essentially the same in the low $\tan\beta$ regime. The charged Higgs-fermion couplings keep their MSSM form, but the mass is shifted 
by a term depending on the value of $\lambda$ \cite{Ellwanger:2009dp}. 
However for the same $m_{H^\pm}$ eigenvalue, independent of its theoretical origin in the two models, the $\Delta F=2$ contributions are in practice identical.

It is convenient to parameterize new physics effects in the $\Delta F=2$ amplitude through the relation,
\begin{equation}
M_{12}^{s(d)} = (M_{12}^{s(d)})_{SM} (1+ h_{s(d)} e^{2 i \sigma_{s(d)}})
\end{equation} 
In standard MFV  ($i.e.,~U(3)^3$),  NP-contributions satisfy $h_d =h_s \equiv h$ while for phases $\sigma_d =\sigma_s =0$. In $U(2)^3$ the same relation holds for $h_{s(d)}$ but  phases are no longer zero, $i.e.,~\sigma_d =\sigma_s \equiv\sigma \ne 0$. In terms of B-meson mass differences in MFV we can therefore express NP-contributions in a universal manner, as
\begin{equation}
|1+h e^{2 i \sigma}| = \frac{\Delta M_{s(d)}}{ \Delta M_{s(d)}^{SM}}
\end{equation}
By considering only the charged Higgs  and 
chargino (higgsino) contributions to $h$, we have
\begin{equation}
h =F_{H^{\pm}} + F_{\tilde H^{\pm}}
\end{equation}
where $F_{H^{\pm}} $ and $ F_{\tilde H^{\pm}} $ stand for the  percentage deviations ($i.e.,~F\times 100\%$) w.r.t the SM prediction, induced by charged Higgs and higgsino contributions, respectively. 

The Higgsino contributions depend, besides $\tan\beta$ and  the 
Higgsino mass $m_{\tilde H^{\pm}}$, on the soft stop mass $m_{U_{33}}$. In hMSSM due to the requirement of very heavy stop masses, these effects decouple. In (MFV-)NMSSM where there is no such requirement such effects are still minor, giving a maximal contribution through $F_{\tilde H^{\pm}}$ of order  $\sim 3\%$\cite{Barbieri:2014tja}. On the other hand, charged Higgs contributions being
mainly a function of $m_{H^{\pm}}$ and $\tan \beta$ are instead significant, inducing large deviations to the SM prediction. To see how these are affected by LHC bounds we turn the results of our previous discussion into constraints on the $m_{H^{\pm}}-\tan\beta $ plane of hMSSM and NMSSM 
\footnote{The Higgs sector of MSSM at tree level
can be described
by only two parameters \emph{i.e,} $m_A(m_{H^{\pm}})$ and $\tan\beta$, but the Higgs
sector of NMSSM involves six free parameters. To obtain the exclusion in the latter case we vary all parameters randomly within the parameter space of \eqref{nmssmpara}
and look for the points
which are excluded in the $m_{H^{\pm}}-\tan \beta$ plane.}. 
Note that in our figures we also take into account the limits from the $125~GeV$ Higgs observables which impose strong additional constraints to the case of NMSSM.
\begin{figure}[t]
\centering
  \includegraphics[width=0.8\textwidth]{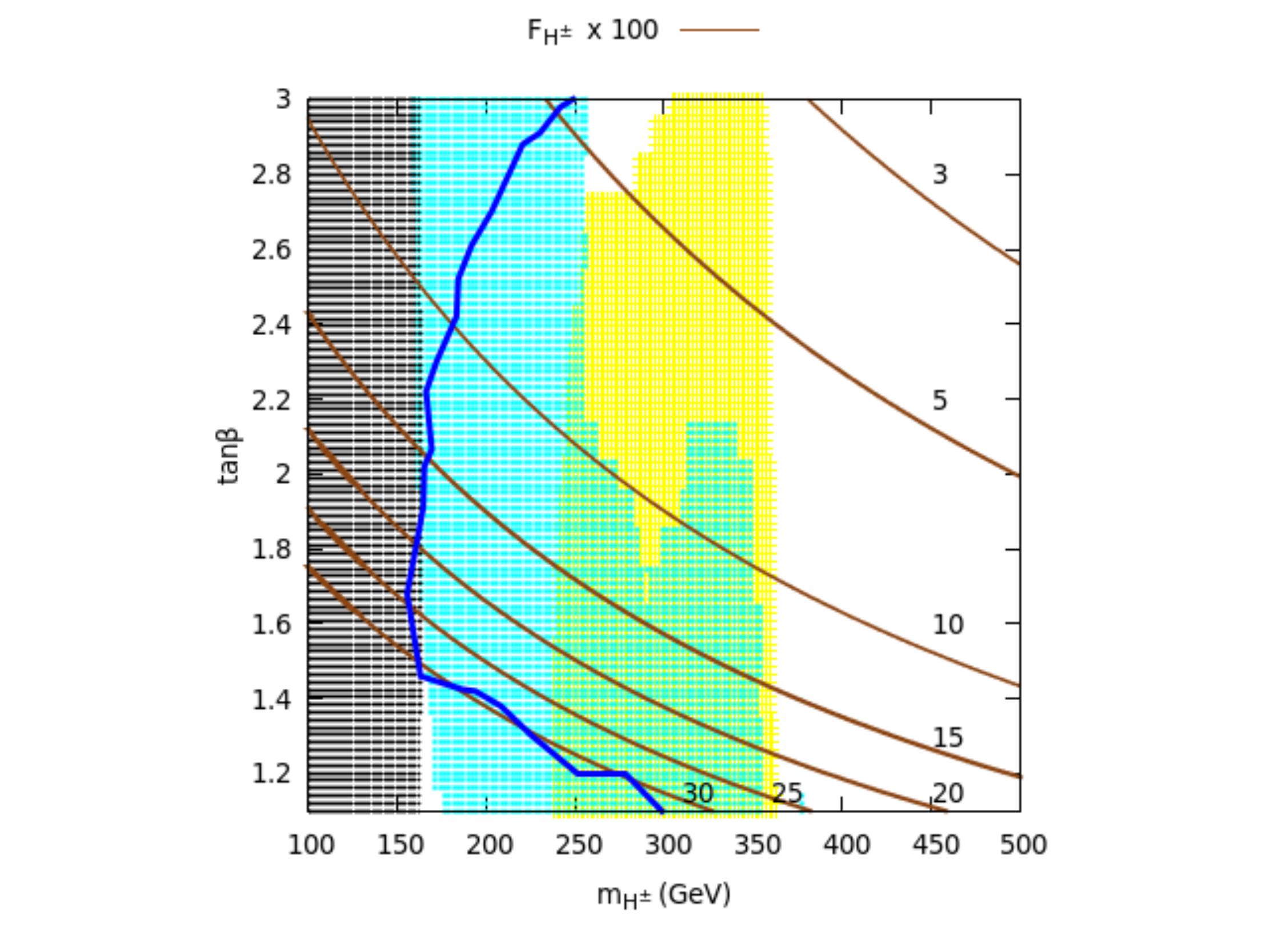} 
\caption{\small Brown contours shows percentage modification 
$F_{H^{\pm}}$ to $\Delta F=2$ observables involving charged Higgs. 
Gray $(H^+ \rightarrow \tau^+ \nu)$, cyan $(H \rightarrow ZZ)$ and yellow $(A \rightarrow hZ)$ regions are hMSSM exclusions at $95 \% CL$. NMSSM exclusion is on the left-side of the blue contour.}
\label{ma-tb-mssm}
\end{figure}\normalsize

In Fig.\ref{ma-tb-mssm} we present the exclusions from direct 
search constraints and Higgs observables together 
with the contours of charged Higgs percentage modification $F_{H^{\pm}}\times 100\%$ to B-meson mass differences, on the $m_{H^\pm} -\tan \beta$ plane. The blue line represents the NMSSM bound, on the left side of which all points are currently excluded\footnote{Note that we do not consider
constraints from $g-2$ and Dark matter on $m_{H^{\pm}}$ and $\tan \beta$
since these are sensitive to other irrelevant parameters  
$e.g., M_1, M_2$, which we have kept fixed.}. The colored regions represent the excluded hMSSM points due to the non-observation of scalars in Heavy-Higgs searches. As shown in colored areas, the region of hMSSM for $m_{H^\pm} \lesssim 350~ GeV$ and  $\tan \beta< 3$ is almost completely excluded.  A close to $\sim 25 \% $ contribution for  $F_{H^{\pm}}$ is still allowed, however this is restricted to a tiny portion of the allowed parameter space with $\tan\beta\lesssim 1.2$ and $m_{H^\pm}: 350-400~GeV$. In NMSSM the relevant constraints are considerably more relaxed, still allowing for a  $\sim 30\%$ effect, even without taking into account the higgsino additional $2-3\%$ contribution, for the case of light stops.

From our previous discussion in Sec.3.1 it is expected that 
the direct search limits alone cannot 
put strong constraints on the $m_{H^\pm} -\tan \beta$ plane of NMSSM, due to the 
diverse patterns of the BR's  \cite{Guchait:2015owa,Kumar:2016tdh}. 
However when the constraints from Higgs observables and LEP 
are also included then a large region for $m_{H^\pm}$ and low $\tan \beta$ becomes excluded. In hMSSM the situation is different and direct searches exclude a larger portion of the parameter space, even without taking into account other constraints. In fact, restrictions from  Higgs observables in this case are eventually found to lie within the already excluded region of parameter space \cite{Djouadi:2013vqa}. For our Fig.\ref{ma-tb-mssm}, we have implemented all these limits in {\tt NMSSMTools} and scanned the parameter space of eq.\eqref{nmssmpara}.


A final remark concerns the future LHC prospects with respect to our results. As has been noted by the authors of \cite{Djouadi:2013vqa} the direct search limits, here applied with respect to flavour physics, cannot be extended to higher scales until the $t \bar t$ and  $t \bar b$ channels, are improved. Since above the top quark threshold and for low $\tan\beta$ the dominant decay modes are $(H,A) \rightarrow t \bar t$ and 
$H^{+} \rightarrow t \bar b $ such experimental measurements will restrict further the allowed charged Higgs masses, at least in the case of hMSSM. These masses, as discussed, essentially control the magnitude of the dominant contributions in $\Delta F=2$ observables within the framework of MFV. Therefore, as our analysis suggests, a potential non-observation of new scalars beyond $\sim 350~GeV$ will also result to a significant suppression of the maximal allowed MFV effects in B-meson mass differences.          

\paragraph{\textbf{\emph{Note added:}}}\footnote{Thanks to Junjie  Cao for pointing out the available public releases and preliminary reports from Atlas and CMS.} \emph{While this work was in preparation
certain preliminary results  at $\sqrt{s}=13~TeV$ have become  available. In particular $H \rightarrow ZZ$ \cite{ATLAS-CONF-2016-082}, $A \rightarrow hZ$ \cite{ATLAS-CONF-2016-015}, 
$H^{\pm} \rightarrow \tau \nu$ \cite{Aaboud:2016dig} and  more importantly the newly updated limits in $H^{\pm} \rightarrow t \bar b$ \cite{ATLAS-CONF-2016-089} channel. For the moment they seem to disfavour a charged higgs mass below $\sim 400 ~ GeV$ for very low $\tan\beta\lesssim 1.5$ and therefore indicate a suppression of maximal-MFV in hMSSM, accordingly (Fig.\ref{ma-tb-mssm}). Once more data is available, especially in $H \rightarrow t \bar t $ and $H^{\pm} \rightarrow t \bar b$,
stronger constraints on the $m_{H^{\pm}}-\tan \beta$ planes are expected.}

\section{Summary and conclusions}\label{4}
We may now briefly summarize the main points of our analysis. The fact that usually NMSSM models do not deviate significantly from their respective MSSM-limits, in their predictions on $\Delta F=2$ processes, motivated us to search for effects that can reverse this typical behaviour. We find for $Z_3$-NMSSM that sizeable genuine-NMSSM contributions  may arise either from certain neutralino-gluino box diagrams or from double penguins both effective for large $\tan\beta$, and under different circumstances. To reverse the typical behaviour of NMSSM effects (being subleading) an enhancement mechanism was expected to take place. Therefore, we studied theoretically those mechanisms by isolating any possible source of genuine-NMSSM effects without considering in advance a specific susy-spectrum or flavour structure. This is what eventually led us to regions of parameter space where such effects were expected to give sizeable contributions and our subsequent numerical analysis reckoned them.

In brief, we mention that neutralino-gluino boxes can contribute significantly when higgsino-singlino ($ \tilde H_d^0 - \tilde S$) mixing is sufficiently large, typically requiring $\lambda\sim k\gtrsim 0.5$ and $\mu_{eff}\lesssim 300~ GeV$ in susy-models with sizeable gluino-gluino contributions. On the other hand, double penguin diagrams require, a light mass for the CP-even or (preferably) CP-odd singlet scalars and a relatively light mass for the heavy Higgs doublets. The latter requirement enhances genuine NMSSM-contributions even for light singlet masses away from the resonance, which is always present at $m_{h(a)}^3\sim M_{B_q}$ for $\Delta M_{q}$ observables. However, it is not easily obtained in $Z_3$-NMSSM due to strong constraints from the Higgs potential setting the heavy Higgs mass typically above $1~TeV$ even for small $\lambda$. The explicit value of $\lambda$ in this case is not directly relevant to the size of the effects as long as it is sufficiently large to distinguish between the NMSSM model and its respective MSSM-limit (\emph{i.e., $\lambda\gtrsim 0.1$})\footnote{The size of $\lambda$  is, however, \emph{indirectly important} since it controls the value of $v_s$ (together with $\mu_{eff}$) and the allowed values of $M_A, Z^{13}_{h(a)}$  through the minimization conditions,  with all these parameters being responsible for the size of genuine-NMSSM effects in Double Penguins, as discussed.}.

In the second part of our study (sec.\ref{3}) we discussed how the LHC Run-I limits from heavy-Higgs non-observation along with Higgs observables can be translated into different bounds in the $m_H^{\pm}-\tan\beta$ planes of MSSM and NMSSM. This, essentially allows to distinguish between the two models in regions where their predictions with respect to $\Delta F=2$ observables are expected to be identical. Thus we have included an analysis of the maximal currently allowed NP-contributions, in MFV models, updating the relevant bounds with these considerations.  

We finally conclude with a general remark on our approach. Our analysis, especially in the first part (sec.\ref{2}), was intended to be both inclusive and exclusive. In this sense, we point out that sizeable genuine NMSSM effects (barring accidental cancellations) are not expected to lie far beyond the parameter space of our analysis. Being always induced by an enhancement mechanism, we arrive at the conclusion that for such effects to give significant contributions elsewhere, another mechanism (which escaped our attention) is expected to underlie.

\section{Acknowledgements}
JK is grateful to David Straub for valuable discussions and for motivating the initial idea of the project. Also wishes to thank Gobinda Majumder for discussions related to CMS and ATLAS limits, Ulrich Ellwanger for correspondence on {\tt NMSSMTools} and Monoranjan Guchait, Charanjit S. Aulakh for discussions and encouragement. JK also thanks Disha Bhatia for useful conversations. MP would like to thank Athanasios Dedes and Janusz Rosiek for helpful discussions. Both authors wish to further express their gratitude to Monoranjan Guchait for proofreading and Janusz Rosiek for proofreading and guidance in various aspects of \code~.

\newpage
\appendix
\renewcommand{\thesection}{Appendix~\Alph{section}}
\renewcommand{\thesubsection}{\Alph{section}.\arabic{subsection}}
\renewcommand{\theequation}{\Alph{section}.\arabic{equation}}
\section{Wilson Coefficients for gluino related box contributions}\label{app:ngwc}
We display here for reference the Wilson Coefficients in mass basis for neutralino-gluino and gluino-gluino box contributions for $B_s$-mixing  in MSSM, taken from \cite{Altmannshofer:2007cs} (consistent with  \cite{Rosiek:1995kg},\code) and transformed to our operator basis. The WC for the NMSSM can
be easily obtained from the MSSM ones by simply extending the 
running of the neutralino index $a$, from 4 to 5 in all relevant summations. The squark indices ($k,l$), run as usual from 1 to 6. For $B_d$ mixing one needs to make the index replacements 
$2 \rightarrow 1$ and $5\to 4$ in all expressions.
\subsection{Neutralino-gluino contributions}
\small
\begin{align}
 C^{VLL} &=-\frac{g_3^2}{16 \pi^2} \frac{1}{6} ~V^L_{2ka}~ 
V^{L*}_{3la} ~Z_{2l} ~Z_{3k}^{*}
~D_2(m_{\tilde g}^2,m_{a}^2,m_k^2,m_l^2) \nonumber\\  
~& - \frac{g_3^2}{16 \pi^2}  \frac{1}{6} ~\left (V^L_{2ka} ~V^L_{2la}
~Z_{3k}^* ~Z_{3l}^* ~+~  V^{L*}_{3ka} ~ V^{L*}_{3la} ~
Z_{2k} ~Z_{2l} ~ \right ) ~ m_{\tilde g} ~ m_a ~ D_0(m_{\tilde g}^2,m_{a}^2,m_k^2,m_l^2) \\
 C^{VRR} &= -\frac{g_3^2}{16 \pi^2}\frac{1}{6} 
~V^R_{2ka}~ V^{R*}_{3la} ~Z_{5l} ~Z_{6k}^{*}
~D_2(m_{\tilde g}^2,m_{a}^2,m_k^2,m_l^2) \nonumber\\  
~& -\frac{g_3^2}{16 \pi^2}   \frac{1}{6} ~\left (V^R_{2ka} ~V^R_{2la}
~Z_{6k}^* ~Z_{6l}^* ~+~   V^{R*}_{3ka} ~ V^{R*}_{3la}~~
Z_{5k} ~Z_{5l} ~
\right ) ~ m_{\tilde g} ~ m_a ~ D_0(m_{\tilde g}^2,m_{a}^2,m_k^2,m_l^2) \\
 C_1^{SLL} & =  \frac{g_3^2}{16 \pi^2}\frac{7}{6}
~V^L_{2ka} ~V^{R*}_{3la} ~Z^*_{6k}~Z_{2l}
~ m_{\tilde g} ~m_a ~D_0(m_{\tilde g}^2,m_{a}^2,m_k^2,m_l^2)\nonumber  \\
 ~ & + \frac{g_3^2}{16 \pi^2} ~  \frac{1}{6}  ~ 
\left (V^{R*}_{3ka} ~V^{R*}_{3la} ~ Z_{2k}~ Z_{2l} ~+~ V^L_{2ka} ~V^L_{2la} 
~Z^*_{6k}~Z^*_{6l}       \right) ~m_{\tilde g} ~m_a ~
D_0(m_{\tilde g}^2,m_{a}^2,m_k^2,m_l^2) \\
 C_{1}^{SRR} & = \frac{g_3^2}{16 \pi^2}\frac{7}{6} 
~V^R_{2ka} ~V^{L*}_{3la} ~Z^*_{3k}  Z_{5l} 
~ m_{\tilde g} ~m_a ~D_0(m_{\tilde g}^2,m_{a}^2,m_k^2,m_l^2) \nonumber \\
 ~ & + \frac{g_3^2}{16 \pi^2} ~  \frac{1}{6} ~ 
\left (V^{L*}_{3ka} ~V^{L*}_{3la} ~ Z_{5k}~ Z_{5l} ~ +~ V^R_{2ka} ~V^R_{2la} 
~Z^*_{3k}~Z^*_{3l}\right)   ~m_{\tilde g} ~m_a ~
D_0(m_{\tilde g}^2,m_{a}^2,m_k^2,m_l^2)      \\
 C_{2}^{SLL} & = -\frac{g_3^2}{16 \pi^2} \frac{1}{24} 
~V^L_{2ka} ~V^{R*}_{3la} ~Z^*_{6k} ~Z_{2l}
~m_{\tilde g} ~m_a  ~D_0(m_{\tilde g}^2,m_{a}^2,m_k^2,m_l^2) \nonumber \\ 
 & + \frac{g_3^2}{16 \pi^2}~ \frac{1}{24}
\left (V^{R*}_{3ka} ~V^{R*}_{3la} ~ Z_{2k} ~Z_{2l} ~
+ ~ V^L_{2ka} ~ V^L_{2la} ~Z^*_{6k} ~ Z^*_{6l}  \right ) ~m_{\tilde g} ~
m_a ~ D_0(m_{\tilde g}^2,m_{a}^2,m_k^2,m_l^2) 
\\
  C_{2}^{SRR} & = -\frac{g_3^2}{16 \pi^2} \frac{1}{24} 
~V^R_{2ka} ~V^{L*}_{3la} ~Z^*_{3k} ~Z_{5l} 
~m_{\tilde g} ~m_a  ~D_0(m_{\tilde g}^2,m_{a}^2,m_k^2,m_l^2)\nonumber  \\ 
 &+ \frac{g_3^2}{16 \pi^2}~ \frac{1}{24} 
\left (V^{L*}_{3ka} ~V^{L*}_{3la} ~ Z_{5k} ~Z_{5l} ~
+ ~ V^R_{2ka} ~ V^R_{2la} ~Z^*_{3k} ~ Z^*_{3l}  \right )~m_{\tilde g} ~
m_a ~ D_0(m_{\tilde g}^2,m_{a}^2,m_k^2,m_l^2)   
\end{align}
\begin{align}
 C^{VLR} & =-\frac{g_3^2}{16 \pi^2} \frac{1}{4} ~\left (
V^R_{2ka} ~V^{R*}_{3la} ~Z_{2l} ~Z_{3k}^{*}  ~
+~ V^L_{2ka} ~V^{L*}_{3la} ~Z_{5l} ~Z_{6k}^*  \right )
~D_2(m_{\tilde g}^2,m_{a}^2,m_k^2,m_l^2)  \nonumber \\
 & +\frac{g_3^2}{16 \pi^2} \frac{1}{6} ~\left (V^R_{2ka} ~V^{L*}_{3la}
~Z_{6k}^* ~Z_{2l} ~+~ ~V^{L}_{2ka}~ V^{R*}_{3la} ~
~ Z_{5l} ~Z_{3k}^*
\right )~m_{\tilde g} ~m_a ~D_0(m_{\tilde g}^2,m_{a}^2,m_k^2,m_l^2) \nonumber \\
 & + \frac{g_3^2}{16 \pi^2} \frac{1}{12}    
  \Big(~3~V^{L*}_{3ka} ~V^{R*}_{3la} ~Z_{2l} ~Z_{5k}
~+~3~ V^L_{2ka} ~V^R_{2la} ~ Z^{*}_{3l} ~Z_{6k}^*  \nonumber \\
 ~&\hspace{1.6cm}+~ V^{L*}_{3la} ~V^{R*}_{3ka} ~ Z_{5k} ~Z_{2l} ~ 
~+~ V^R_{2ka} ~V^L_{2la} ~Z^*_{3l} ~Z_{6k}^*\Big)~D_2(m_{\tilde g}^2,m_{a}^2,m_k^2,m_l^2)  \\ \nonumber
\\
 C^{SLR} &= -\frac{g_3^2}{16 \pi^2} \frac{1}{6}~
(V^R_{2ka} ~ V^{R*}_{3la}  ~ Z_{2l} Z_{3k}^* ~+~ V^L_{2ka} ~
~ V^{L*}_{3la} ~ Z_{5l} ~ Z^*_{6k}) ~
D_2(m_{\tilde g}^2,m_{a}^2,m_k^2,m_l^2)  \nonumber \\
 ~& + \frac{g_3^2}{16 \pi^2}
~(V^R_{2ka} ~V^{L*}_{3la}  ~Z^*_{6k} Z_{2l} ~+~ V^L_{2ka}~ 
~V^{R*}_{3la} ~Z_{5l} Z^*_{3k} )~ m_{\tilde g} 
~m_a~ D_0(m_{\tilde g}^2,m_{a}^2,m_k^2,m_l^2)  \nonumber \\
 ~& +\frac{g_3^2}{16 \pi^2}~ \frac{1}{6} 
~\Big(V^{L*}_{3ka} ~V^{R*}_{3la} ~Z_{2l} ~Z_{5k} ~
+~  V^L_{2ka}~ V^R_{2la}~ Z^*_{3l} ~Z^*_{6k} ~  \nonumber \\
 &\hspace{1.6cm}+~ 3 ~V^{L*}_{3la} ~V^{R*}_{3ka} ~Z_{5k}~ Z_{2l} 
~+~3~ V^R_{2ka}~ V^L_{2la} ~Z^*_{3l} ~Z^*_{6k}  ~ \Big)~ D_2(m_{\tilde g}^2,m_{a}^2,m_k^2,m_l^2) 
\end{align}

\subsection{Gluino-gluino contributions}
\begin{align}
 C^{VLL} &= -\frac{g_3^4}{16 \pi^2}\frac{1}{36} ~Z_{2k} Z_{2l} 
~Z_{3k}^* ~Z_{3l}^{*}
~\Big(11~D_2(m_{\tilde g}^2,m_{\tilde g}^2,m_k^2,m_l^2)   
 ~+ 4 m_{\tilde g}^2 ~D_0(m_{\tilde g}^2,m_{\tilde g}^2,m_k^2,m_l^2)\Big) \\
 C^{VRR} &=-\frac{g_3^4}{16 \pi^2} \frac{1}{36} ~Z_{5k} Z_{5l} 
~Z_{6k}^* ~Z_{6l}^{*}
~\Big(11~D_2(m_{\tilde g}^2,m_{\tilde g}^2,m_k^2,m_l^2)   
 ~+ 4 m_{\tilde g}^2 ~D_0(m_{\tilde g}^2,m_{\tilde g}^2,m_k^2,m_l^2)\Big) \\
C_{1}^{SLL} & =-\frac{g_3^4}{16 \pi^2} \frac{37}{36} ~Z_{2k} Z_{2l} ~
Z_{6k}^* Z_{6l}^{*}~ m_{\tilde g}^2 
~D_0(m_{\tilde g}^2,m_{\tilde g}^2,m_k^2,m_l^2) \\ \nonumber
\\  
C_{1}^{SRR} & =-\frac{g_3^4}{16 \pi^2} \frac{37}{36} ~Z_{5k} Z_{5l} ~
Z_{3k}^* Z_{3l}^{*}~ m_{\tilde g}^2 
~D_0(m_{\tilde g}^2,m_{\tilde g}^2,m_k^2,m_l^2)  \\
\nonumber  \\
C_{2}^{SLL} & = \frac{g_3^4}{16 \pi^2} \frac{1}{48}~ Z_{2k} Z_{2l}~ 
Z_{6k}^* Z_{6l}^{*}~ m_{\tilde g}^2 
~D_0(m_{\tilde g}^2,m_{\tilde g}^2,m_k^2,m_l^2)  \\
\nonumber \\
 C_{2}^{SRR} & =\frac{g_3^4}{16 \pi^2} \frac{1}{48}~ Z_{5k} Z_{5l}~ 
Z_{3k}^* Z_{3l}^{*}~ m_{\tilde g}^2 
~D_0(m_{\tilde g}^2,m_{\tilde g}^2,m_k^2,m_l^2)  \\
\nonumber 
\\
C^{VLR} & = \frac{g_3^4}{16 \pi^2} \frac{1}{18} ~Z_{2k} Z_{5l} ~
Z_{3k}^* Z_{6l}^{*}  ~m_{\tilde g}^2 ~
D_0(m_{\tilde g}^2,m_{\tilde g}^2,m_k^2,m_l^2) \nonumber \\
 &- \frac{g_3^4}{16 \pi^2} \frac{5}{36} ~Z_{2k} Z_{5l} 
~(3~ Z_{6k}^* ~Z_{3l}^{*} ~-~2~ Z_{3k}^* Z_{6l}^{*} )
~D_2(m_{\tilde g}^2,m_{\tilde g}^2,m_k^2,m_l^2) 
\end{align}
\begin{align}
 C^{SLR} & =-\frac{g_3^4}{16 \pi^2} \frac{7}{3} ~Z_{2k} Z_{5l} 
~Z_{3k}^* Z_{6l}^{*}~  m_{\tilde g}^2 ~
D_0(m_{\tilde g}^2,m_{\tilde g}^2,m_k^2,m_l^2) \nonumber \\
 & +\frac{g_3^4}{16 \pi^2} \frac{1}{18} ~Z_{2k} Z_{5l} 
~(6~ Z_{3k}^* ~Z_{6l}^{*} ~+~11 ~Z_{6k}^* ~Z_{3l}^{*})
~D_2(m_{\tilde g}^2,m_{\tilde g}^2,m_k^2,m_l^2) 
\end{align}
\section{Loop Functions for zero external momenta}\label{app:loop}
\small
\begin{align}
C_0 (m_1^2, m_2^2, m_3^2) & = -\Big( \frac{m_1^2 \log m_1^2 }{(m_3^2-m_1^2)(m_2^2-m_1^2)}
+ ~(1 \leftrightarrow 2)~ + ~(1 \leftrightarrow 3)\Big) \\
\nonumber \\
D_0 (m_1^2, m_2^2, m_3^2, m_4^2) & = \frac{m_1^2 \log m_1^2 }{(m_4^2-m_1^2)(m_3^2-m_1^2)(m_2^2-m_1^2)}
+ ~(1 \leftrightarrow 2)~ + ~(1 \leftrightarrow 3) + ~(1 \leftrightarrow 4) \\
\nonumber \\
E_0(m_1^2, m_2^2, m_3^2, m_4^2,m_5^2)& = -\Big(\frac{m_1^2 \log m_1^2 }{(m_5^2-m_1^2)(m_4^2-m_1^2)(m_3^2-m_1^2)(m_2^2-m_1^2)}
+ ~\dots ~ + ~(1 \leftrightarrow 5)\Big)  \\
D_2 (m_1^2, m_2^2, m_3^2, m_4^2) & = \frac{m_1^4 \log m_1^2 }{(m_4^2-m_1^2)(m_3^4-m_1^2)(m_2^2-m_1^2)}
+ ~(1 \leftrightarrow 2)~ + ~(1 \leftrightarrow 3) + ~(1 \leftrightarrow 4) 
\end{align}
\normalsize

\section{Minimization of NMSSM potential for large  $\lambda$, $\tan\beta$}\label{app:hmin}

Here is discussed a certain analytical method for obtaining phenomenologically viable minimization conditions in the Higgs potential of NMSSM, which is also effective in the large $\lambda$-$\tan\beta$ region. Various aspects of this tree-level approach have been discussed in the past \cite{Ellwanger:2006rm,Miller:2003ay}. In this method a large $M_A$ and together with it large soft masses are required, raising questions related to naturalness and fine-tuning. However, we find that it gives the most natural tree-level minimization solution in the large $\tan\beta,\lambda$ regime. Moreover, the Barbieri-Giudice (BG) fine-tuning parameters \cite{Barbieri:1987fn} as calculated from \texttt{NMSSMTools} and taking into account loop corrections, indicate effectively zero additional fine-tuning from the genuine-NMSSM parameter space. In particular, the typical range of the BG measure in the parameter space of sec.\ref{2} is $\Delta_{\max}\sim 4-10$, driven by the MSSM soft masses $m_{H_u},m_{H_d}$ and with the NMSSM sector giving  the maximal effect through $\Delta_{A_\lambda}\sim 4$.

In light of Higgs observables from LHC, supporting a SM-like Higgs particle at $125\, GeV $, this method acquires a renewed interest. This is because, as will be discussed, it essentially decouples  all other CP-even states (each in a different sense) from the SM-like Higgs boson, independent of the explicit value of the $\lambda,\kappa$ parameters. As a result, the Higgs observables remain practically SM-like and a consistency with the Higgs phenomenological constraints (or other) is always present in our \texttt{NMSSMTools} scans. 

In what follows we present the relations obtained in this approach and then examine their ``naturalness" with respect to the tree level minimization conditions.  Only in this appendix, we switch to the conventions of \cite{Ellwanger:2009dp} for convenience to the reader, using a more familiar notation commonly used in NMSSM studies. This is easily achieved (for expressions shown here) by simply redefining 
\begin{equation}
(\bar v_u ,\bar v_d,s) \equiv {1\over \sqrt{2}} (v_u, v_d, v_s)  
\end{equation}
corresponding to a different convention in the definition of all vevs.  Long analytical expressions which are not directly relevant to this approach (but could give a more self-consistent description), are neglected\footnote{They can be taken directly from our references.}. Real parameters are considered for simplicity while the convention where $\bar v_u,\bar v_d,(\kappa s)$  are positive, is followed. Finally, the reader should always bear in mind that also here, we refer to a generalized concept of ``flavour", as explained in introduction.

First, we discuss the CP-even sector of NMSSM. As well known, one can use the minimization conditions to eliminate the dependence on all soft squared masses and therefore there are only six, free parameters in the $Z_3$-invariant potential of NMSSM, namely $\lambda ,\kappa ,\mu_{eff}({ \,or\, } s),$ $ A_\lambda ({\, or\, } B_{eff}), A_\kappa,\tan\beta$. The Higgs mass matrix at tree level, in the initial basis $(H_d,H_u,S)$  reads,
$$\bm{M}^2_H=\left(\begin{array}{ccc}
 {g^2 \bar v_d^2} +\mu B\tan\beta    &(2\lambda^2-g^2) \bar v_u \bar v_d-\mu B & \lambda \bar v_d\big(2 \mu-(B+\kappa s)\tan\beta\big) \\[2mm] 
• & g^2 \bar v_u^2 +{\mu B \over\tan\beta} &\lambda \bar v_d\big(2 \mu\tan\beta-(B+\kappa s)\big) \\ [2mm]
• & • &\lambda A_\lambda  {\bar v_u \bar v_d\over s}+\kappa s\big(A_\kappa+4 \kappa s\big)
\end{array}\right) $$
where following \cite{Ellwanger:2009dp}, we denote 
\begin{eqnarray*}
&\mu\equiv\mu_{eff}\,,~s\equiv <S>= \mu_{eff}/\lambda\,,~B\equiv B_{eff}=A_\lambda +\kappa s\\
&\bar v^2 = \bar v_u^2+\bar v_d^2\simeq(174 ~GeV)^2\, ,~ g^2\equiv {g_1^2 +g_2^2\over 2}\,,~ M_Z^2=g^2 \bar v^2\,,~ M_A^2\equiv{2\mu B\over\sin{2\beta}}
\end{eqnarray*} 

One can bring ${M}_H^2$ to a more convenient form by rotating with a $\tan\beta$-related $2\times 2$ block-rotation  matrix which mixes only the doublet states $(H_d,H_u)$. The relevant orthogonal matrix is parameterized as
$$\bm{R}(\beta)=\left(\begin{array}{ccc}
\sin\beta &-\cos\beta &\\
\cos\beta &\sin\beta&\\
 && 1
\end{array}\right) $$

  In this rotated flavour (gauge) eigenstate basis\footnote{This is obtained in our notation as $(\hat H_d,\hat H_u,S)^{\top}= \bm{R}(\beta)~ ( H_d,H_u,S)^\top $. Notice that for vary large $\tan\beta$, the mixing induced here is suppressed since $\bm{R}\simeq \bm{I}$. Our hatted notation should not be confused with the hatted superfield notation used in text} $(\hat H_d,\hat H_u,S)$ the mixing of the new states is controlled by the off-diagonal elements,
\begin{align*}
&( {\hat M}^2_H)_{12}= \big( M_Z^2-\lambda^2 \bar v^2  \big) \cos{2 \beta} \sin{2 \beta}\\
&( {\hat M}^2_H)_{13}= \lambda \bar v (B+\kappa s) \cos{2\beta}\\
&( {\hat M}^2_H)_{23}=  \lambda \bar v\big(2\mu -(B + \kappa s)\sin{2\beta}\big)
\end{align*}
while the diagonal elements read,
\begin{align*}
&( {\hat M}^2_H)_{11}= M_A^2+ (M_Z^2-\lambda^2 \bar v^2)\sin^2{2\beta} \\
&( {\hat M}^2_H)_{22}= M_Z^2 \cos^2{2\beta}+\lambda^2 \bar v^2\sin^2{2\beta}  \\
&( {\hat M}^2_H)_{33}=( {M}^2_H)_{33}.
\end{align*}

For $M_A>M_Z$ and large $\tan\beta$, the lighter doublet state in this new basis is $\hat H_u$ with a squared ``flavour" mass $( {\hat M}^2_H)_{22}\simeq M^2_Z$. When the mixing with the other CP-even states is negligible then $\hat H_u$ dominates in the  SM-like Higgs mass eigenstate. In the case where $\hat H_u$ is the lightest state then any mixing with the other flavour eigenstates can only lead to a lighter mass eigenvalue for the SM-like Higgs. If instead the singlet is the lightest state, then $\hat{H}_u$  still dominates the SM-Higgs but now it is primarily related to the second lightest mass eigenvalue. This mass can then exceed $M_Z$ at tree level but only at the cost of $\hat H_u-S$ mixing, which is a situation we wish to avoid for phenomenological reasons\footnote{At large $\tan\beta$ this mixing is additionally constrained by the minimization conditions and can easily lead to a tachyonic spectrum when it is large}. Irrespective of the explicit mass hierarchy one simply requires that the mixing of $\hat H_u$ with any other CP-even state is suppressed, thus $M_Z$ becomes an absolute upper bound for the tree level SM-like Higgs mass, as in MSSM.

The $\hat H_u-S$ mixing becomes suppressed, when  one requires, 
\begin{equation}
A_\lambda \simeq {2\mu\over \sin{2\beta}}-2\kappa s= 2\mu \Big({1\over \sin{2\beta}}-{\kappa\over\lambda}\Big)\label{eq:Alam}
\end{equation}
which makes $(\hat M_H^2)_{23}$ small by assumption. On the other hand $\hat H_u-\hat H_d$ mixing is already suppressed at tree level for large $\tan\beta$ due to the presence of a $\sin{2\beta}$ factor in $(\hat M_H^2)_{12}$. In addition, the doublet flavour masses are expected to display a large hierarchy of the form $( {\hat M}^2_H)_{11}\gg ( {\hat M}^2_H)_{22}$ which further suppresses the doublet mixing, in a manner analogous to the decoupling limit of 2HDM. This is understood from eq.\eqref{eq:Alam} which suggests that the natural scale for $M_A$ at large $\tan\beta$ and $\lambda\sim\kappa$ is $M_A\sim A_\lambda \sim \mu \tan\beta$. Thus,  
$$({\hat M}^2_H)_{11} /  ({\hat M}^2_H)_{22} \simeq M_A^2 / M_Z^2 \sim (\mu^2 / M_Z^2)\tan^2{\beta}.$$
Therefore in this approach the heavy doublet decouples from the light one in the usual sense, due to its large mass. The decoupling of the singlet state from the SM-like Higgs has been instead obtained by suppressing the relevant mixing through eq.\eqref{eq:Alam}. The latter method resembles the alignment limit of 2HDM, where the mixing is suppressed by assumption, although here it is applied for light doublet-singlet mixing, only. 

The $\hat H_d-S$ mixing is not directly relevant to the SM-like Higgs doublet state, nevertheless it is associated with the lightest eigenvalue. In this sense it is directly relevant to the consistency and the phenomenological viability of the mass spectrum. Before examining the relevant bounds obtained from the Higgs mass matrices, it is instructive to discuss the asymptotic behaviour of the NMSSM Higgs potential, at large $s$. As has been noted in \cite{Ellwanger:2009dp} for very large values of $s$, the $Z_3$-invariant Higgs potential becomes,
$$V_{Higgs}\sim m_S^2 s_{\infty}^2 +{2\over 3}\kappa A_\kappa s_{\infty}^3 +\kappa^2 s_{\infty}^4$$
The  minimization conditions  are obtained when both 
\begin{align}
s\simeq -{{1\over 4\kappa}} \big( A_\kappa\pm\sqrt{A_\kappa^2-8 m_S^2}\big),
 \\
0\lesssim ~\kappa s~\big(A_\kappa  +4\kappa s\big),~~~~~
\end{align}
are simultaneously satisfied. Requiring the \emph{global minimum} to be located in the range of our convention (\emph{i.e., $\kappa s>0$ under assumption}) one finds $s\simeq {{1\over 4\kappa}} \big( |A_\kappa|+\sqrt{A_\kappa^2-8 m_S^2}\big)$ for $m_S^2<0$ and 
 \begin{equation} 
~~~~~~~~-4\kappa s~ \lesssim A_\kappa ~ \lesssim 0~,~~~(s\equiv \mu / \lambda).\label{eq:Akap}
\end{equation}
In order for this global minimum not to be overtaken by the symmetric ($s=0$) vacuum in the case $m_S^2 > 0$, one needs to impose a stronger upper bound in \eqref{eq:Akap}, namely $A_k\lesssim -3 |m_S|$. Solutions of the asymptotic potential, obtained in the $\kappa s<0$ convention are always symmetric, located at $s'=-s$ for $A'_\kappa= -A_\kappa$, thus they can be produced by a reflection to those discussed here.  The allowed range of \eqref{eq:Akap} is typically valid even for $s$ close to the electroweak scale where terms (linear to $s$) neglected in the asymptotic solution, are expected to be important. On the other hand, when these bounds are violated, negative squared masses in the Higgs sector in general appear. This property is actually expected, since successful minimization is intimately connected to the absence of tachyonic particles in the CP-even and CP-odd sectors of the theory.  

In order to fit a phenomenologically suitable value for $A_\kappa$,  the CP-odd mass matrix needs to be considered, as well\footnote{It can always be taken from \cite{Ellwanger:2009dp} if required.}. One starts in the initial flavour basis $(A_d,A_u,A_s)$ and rotates as previously with ${R}(\beta)$, which also rotates away the neutral Goldstone mode. In the new flavour basis $(\hat A_d, G , A_s)$, we suppress the null Goldstone space and express the remaining CP-odd $2\times 2$ mass matrix through,
\begin{align*}
&( {\hat M}^2_A)_{11}= M_A^2\\
&( {\hat M}^2_A)_{12}= \lambda   (B-3 \kappa s) \bar v\\
&( {\hat M}^2_A)_{22}= \lambda \big( B + 3\kappa s\big){\bar v_u \bar v_d\over s}-3 A_\kappa \kappa s
\end{align*}
The negative contribution in $(\hat M_A^2)_{22}$, obtained for $A_\kappa>0$, essentially drives the CP-odd singlet to negative eigenvalues, a behaviour which becomes considerably worse for large values of $\tan\beta$\footnote{One can  see this from the corresponding determinant  where for $A_\kappa>0$ the relevant (negative) contribution is enhanced by $\tan\beta$ w.r.t surviving terms. Since $M_A^2>0$, a negative determinant signals a negative mass eigenvalue in the CP-odd mass matrix.}. Conversely, by departing from zero through $A_\kappa<0$ the CP-odd singlet acquires rapidly large positive masses with an upper bound set by the $(\hat M_A^2)_{22}$, assuming $M_A^2$ being heavier. The value of $A_\kappa$ also controls the singlet CP-even state through $(\hat M_H^2)_{33}$ which for analogous reasons becomes tachyonic when the lower bound of \eqref{eq:Akap} is violated. Nevertheless, as long as one avoids values close to the edges of \eqref{eq:Akap}, the corresponding mass eigenvalues, related to the two singlets, stay above $\sim 100\, GeV$. One can always consider the central value for reference, namely,
\begin{equation}
A_\kappa\simeq -2 \kappa s .\nonumber \label{eq:Akap2}
\end{equation}

Having described the method of obtaining phenomenologically viable CP-odd and CP-even  masses at tree level, by fitting $A_{\lambda}$ and obtaining the allowed range for $A_\kappa$, we now revisit explicitly the minimization conditions of the $Z_3$-invariant potential. These will give  an insight on the tuning imposed by the conditions \eqref{eq:Alam},\eqref{eq:Akap}. Since  a simultaneous solution to the three minimization conditions at large $\tan\beta$  is required, it is instructive to parameterize them suitably, through 
\begin{align}
\tan\beta &=   {\mu B\over m_{H_u}^2 +\mu^2 + {1\over 2} M_Z^2+(\lambda^2-g^2){\bar v_d^2}}\label{eq:minmhu}\\\nonumber\\
\tan\beta &=  {m_{H_d}^2 +\mu^2+{1\over 2} M_Z^2+(\lambda^2-g^2){\bar v_u^2}\over  \mu B}\label{eq:minmhd}\\ \nonumber\\
\lambda \bar v_u \bar v_d A_\lambda& =  s~\big(m_S^2+\kappa A_\kappa s +2\kappa^2 s^2+\lambda^2 \bar v^2-2\lambda \kappa \bar v_u \bar v_d\big)\label{eq:minms}.
\end{align}
 Next, one can assume a certain hierarchy for the scales involved, which reads 
$$  \big(m_{H_u}^2+\mu^2 +{1\over 2} M_Z^2\big) \tan^2{\beta}  \simeq \mu B \tan\beta\simeq  m_{H_d}^2 $$ 
satisfying identically eq.\eqref{eq:minmhu}\eqref{eq:minmhd} for large $\tan\beta$ without requiring any fine-tuned cancellation between the parameters. Under this assumption, $\mu$ can be taken roughly at a close-to-electroweak scale (denoted also as $\mu$ from now on). The hierarchical condition now turns into
\begin{align*}
\mu ~~~&:~ \sim m_{H_u} \\
 \mu\tan\beta &:~ \sim B\sim m_{H_d} 
\end{align*}
where the splitting of the two scales in controlled only by $\tan\beta$. A natural solution for eq.\eqref{eq:minms} is then obtained at the electroweak level, namely for $s\sim \mu$. In this case, all leading terms contributing to this equation are of electroweak order. Note that the only ``unnatural" parameter appearing in $s$-minimization is $A_\lambda\simeq B \sim \mu\tan\beta$, which however comes together with a $\tan\beta$ suppression through $\bar v_d$.  Therefore, the hierarchical conditions for dimensionful NMSSM parameters, generalize into
\begin{align}
\mu~~~ &:~ \sim m_{H_u}\sim  m_{S} \sim  A_\kappa  ~~~~(\sim s)\nonumber \\
\mu\tan\beta &:~  \sim A_\lambda \sim m_{H_d}  \label{eq:hier}
\end{align} 
while the dimensionless parameters $\lambda,\kappa$,  in this approach remain unconstrained by any reasonable\footnote{An exception to this statement can in principle come for extreme values of $\lambda,\kappa$ which could induce a new hierarchy in the minimization conditions through $\lambda/\kappa \to \tan\beta, \cot\beta  $.} consideration. For large values of $\tan\beta$, 
\begin{equation}
M_A^2 = {2\mu B\over \sin{2\beta}}\simeq \mu A_\lambda \tan\beta\sim (\mu\tan\beta)^2\label{eq:MA}
\end{equation}
is again obtained,
although now through the minimization considerations. This mass scale has been characterized as the ``natural" scale for the heavy Higgs masses, in the past \cite{Miller:2003ay}. It can be easily checked that \eqref{eq:hier} and obviously \eqref{eq:MA} are in agreement with the \emph{fundamental relations} of this method, namely \eqref{eq:Alam},\eqref{eq:Akap}.

\bibliography{deltaf2_v2.bib}
\bibliographystyle{utphys.bst}

\end{document}